\documentclass[aps,prxq,reprint,twocolumn,showpacs,superscriptaddress]{revtex4-2}

\usepackage[utf8]{inputenc}
\usepackage[english]{babel}  

\usepackage{times}
\usepackage{amssymb,amsmath,amsfonts,amsthm}
\usepackage{mathtools}
\usepackage{verbatim}
\usepackage{enumerate}
\usepackage{multirow}
\usepackage{graphicx}
\graphicspath{{pics/}}
\usepackage[colorlinks=true,linkcolor=blue,citecolor=blue,urlcolor=blue]{hyperref}
\addto\captionsenglish{%
  \renewcommand{\figurename}{FIG.}%
}

\makeatletter
\AtBeginDocument{%
  \def\fnum@figure{\figurename~\thefigure}%
}
\makeatother

\usepackage{bbm}
\usepackage{booktabs}

\usepackage{subfigure}

\usepackage{color}
\definecolor{dred}{rgb}{.8,0.2,.2}
\definecolor{ddred}{rgb}{.8,0.5,.5}
\definecolor{dblue}{rgb}{.2,0.2,.8}
\definecolor{dgreen}{rgb}{.2,0.5,.2}

\usepackage{CJKutf8}

\usepackage{array}
\newcolumntype{C}[1]{>{\centering\arraybackslash}p{#1}}
\usepackage{float}

\theoremstyle{plain}

\theoremstyle{definition}

\newcommand{\bra}[1]{\mbox{$\langle #1|$}}
\newcommand{\ket}[1]{\ensuremath{|#1\rangle}}

\newcommand{\be}{\begin{equation}}
	\newcommand{\ee}{\end{equation}}

\usepackage{tikz}
\usepackage{braids}
\usepackage{xifthen}

\usepackage{euscript}
\usepackage{mathrsfs}

\begin{document}

\title{Unveiling the Action of Randomized Compiling on Quantum States: A Geometric Perspective}

\author{Kai Tang}
\affiliation{State Key Laboratory of Quantum Functional Materials, Department of Physics, and Guangdong Basic Research Center of Excellence for Quantum Science, Southern University of Science and Technology (SUSTech), Shenzhen 518055, China}

\author{Chao Zhan}
\affiliation{State Key Laboratory of Quantum Functional Materials, Department of Physics, and Guangdong Basic Research Center of Excellence for Quantum Science, Southern University of Science and Technology (SUSTech), Shenzhen 518055, China}

\author{Yao Song}
\affiliation{Shenzhen Institute for Quantum Science and Engineering (SIQSE), Southern University of Science and Technology, Shenzhen 518055, China}
\affiliation{International Quantum Academy, Shenzhen 518048, China}

\author{Peiling Huang}
\affiliation{Shenzhen Institute for Quantum Science and Engineering (SIQSE), Southern University of Science and Technology, Shenzhen 518055, China}
\affiliation{Department of Physics, Southern University of Science and Technology, Shenzhen 518055, China}

\author{Yuxuan Zhou}
\affiliation{International Quantum Academy, Shenzhen 518048, China}

\author{Song Liu}
\affiliation{International Quantum Academy, Shenzhen 518048, China}

\author{\mbox{Junkai Zeng}}
\affiliation{International Quantum Academy, Shenzhen 518048, China}

\author{Jiansheng Wu}
\affiliation{Shenzhen Institute for Quantum Science and Engineering (SIQSE), Southern University of Science and Technology, Shenzhen 518055, China}

\author{Tongxing Yan}
\affiliation{International Quantum Academy, Shenzhen 518048, China}
    
\author{Xiu-Hao Deng}
\affiliation{International Quantum Academy, Shenzhen 518048, China}
\affiliation{Shenzhen Branch, Hefei National Laboratory, Shenzhen 518048, China}

\author{Yuanzhen Chen}
\email{chenyz@sustech.edu.cn}
\affiliation{State Key Laboratory of Quantum Functional Materials, Department of Physics, and Guangdong Basic Research Center of Excellence for Quantum Science, Southern University of Science and Technology (SUSTech), Shenzhen 518055, China}

\date{\today}

\begin{abstract}

Randomized compiling (RC) converts coherent errors into stochastic Pauli noise, but this conversion alone does not imply isotropic contraction of the output state. Here we use the generalized Bloch vector (GBV) representation to examine the cumulative effect of RC on state purity and orientation in multi-layer circuits. For the independent, isotropically distributed weak coherent errors considered here, the accumulated ensemble-averaged RC channel lies close to the depolarizing direction, producing a state-level near-isotropic contraction that sacrifices a small amount of purity to achieve a substantial suppression of orientational deviation. When dephasing coexists, RC suppresses the coherent-error contribution but leaves a residual deviation dominated by dephasing, setting a limit that we term the ``dephasing wall.'' We experimentally validate these findings on a superconducting quantum processor through GHZ state preparation and phase estimation in a simulation of topological braiding. The results confirm that RC's effectiveness stems from its ability to suppress the orientational bias of coherent errors—a structural transformation not resolved by fidelity alone but directly visible through GBV diagnostics.

\end{abstract}
\maketitle

\section{Introduction}

Reliable quantum computation hinges on effective error management. Various quantum error correction (QEC) and mitigation strategies have been developed to this end~\cite{shor1995scheme,steane1996error,terhal2015quantum,google2025quantum,cai2023quantum,kim2023evidence,quek2024exponentially,wallman2016noise}. The efficacy of such strategies depends on a detailed understanding of the error's nature—not only its magnitude, but also its structural composition. A particularly important structural distinction lies between coherent errors—systematic unitary imperfections arising from control inaccuracies, residual couplings, and crosstalk~\cite{sanders2016bounding,sarovar2020detecting}—and incoherent errors such as decoherence~\cite{wallman2015estimating, kueng2016comparing,katabarwa2015logical}.

The distinct impacts of these error types on a quantum state can be understood through their effects on two fundamental properties of the state: its orientation in state space and its purity \cite{byrd2003characterization,gamel2012measures,tripathi2025benchmarking}. Coherent errors induce pure orientational deviation—systematic rotation of the state at fixed purity \cite{loubenets2020specifying,carignan2024estimating,kaufmann2025characterization}. Incoherent errors generally induce coupled deformation: both purity loss and orientational deviation, with the relative magnitude depending on the specific noise channel. In the special case of depolarizing noise, the deformation becomes isotropic contraction—purity loss without orientational bias \cite{burrell2009geometry}. Unlike their incoherent counterparts, coherent errors can constructively accumulate, leading to a worst-case scaling of algorithmic infidelity that is quadratic in both the error strength and the circuit depth~\cite{magesan2012characterizing,wallman2015estimating, kueng2016comparing}. Such accumulation makes coherent errors particularly detrimental for deep circuits. Tailoring an effective defensive response often requires knowing which type of error dominates, or how they intermix.

Among the various error mitigation tools, randomized compiling (RC) stands out as a particularly effective technique~\cite{wallman2016noise, winick2022concepts,Hashim2021Randomized, hashim2023benchmarking, perrin2024mitigating, goss2024extending, kurita2023synergetic, beale2023randomized, jain2023improved,urbanek2021mitigating,gu2023noise}. At the channel level, it is well understood that RC converts coherent errors into stochastic Pauli errors, thereby improving metrics such as trace distance~\cite{jozsa1994fidelity,gilchrist2005distance,blume2025quantum} and worst-case fidelity~\cite{Hashim2021Randomized,jain2023improved}. Pauli noise is generally anisotropic, however, so this conversion alone does not establish whether the accumulated noise approximately preserves the orientation of the output state. Understanding the structure of the residual error is relevant to deep-circuit performance, algorithm-specific optimization, and the use of RC in QEC~\cite{beale2023randomized,jain2023improved,beale2018coherence}.

Here we employ the generalized Bloch vector (GBV) representation~\cite{kimura2003bloch,loubenets2021bloch} to provide a geometric picture of RC's action on quantum states. In this language, a quantum state is represented by a vector in a Euclidean space: its length encodes purity and its direction naturally defines an orientation for the state~\cite{kimura2003bloch,loubenets2021bloch,bengtsson2017geometry}. Different error types produce distinct geometric signatures: coherent errors rotate the GBV at fixed length; depolarizing noise contracts it isotropically; general incoherent errors do both~\cite{byrd2011general,burrell2009geometry,kimmel2014robust}. This decomposition of orientation and purity makes RC's state-level action directly visible. 

Using the GBV framework, we uncover a cumulative effect of RC in multi-layer circuits that is not captured by the usual single-layer Pauli-noise description. For the independent, isotropically distributed weak coherent errors considered here, the accumulated ensemble-averaged RC channel lies close to the depolarizing direction; correspondingly, the output states undergo a near-isotropic contraction, sacrificing a small amount of purity in exchange for a significant suppression of orientational deviation. In more realistic settings where dephasing coexists with coherent errors, RC suppresses the coherent contribution while leaving dephasing intact. In the regime studied here, the residual orientational deviation is dominated by dephasing, establishing the limit that we term the ``dephasing wall.''

We experimentally verify this geometric picture on a superconducting quantum processor through two complementary tasks. In multi-qubit GHZ state preparation, GBV diagnostics reveal how RC suppresses coherent orientational deviation even under strong decoherence. In a phase estimation task simulating the $D(\mathbb{Z}_3)$ quantum double model, we demonstrate that RC's orientational protection translates directly into algorithmic accuracy and confirm the existence of the dephasing wall. 

The remainder of this paper is organized as follows. Section~\ref{section GBV representation} discusses the GBV representation and its application in numerical simulations of RC. Section~\ref{section GHZ experiment} presents the GHZ state preparation experiment, where GBV diagnostics quantify the purity-orientation trade-off under realistic noise. Section~\ref{section braiding experiment} examines the phase estimation task, showing that orientational protection correlates with algorithmic accuracy. Finally, Section~\ref{section discussion outlook} concludes the paper.

\section{RC studied in the GBV representation \label{section GBV representation}}

In this section, we first formalize the GBV representation and then use numerical simulations to show that RC produces a near-isotropic state-level contraction in multi-layer circuits. We also introduce block-RC as a reduced-overhead variant and specify its range of applicability.

\subsection{The GBV representation \label{GBV representation}}

The density matrix of a \(d\)-dimensional quantum state (for an \(N\)-qubit system, \(d=2^N\)) can be written as
\begin{equation}
\rho = \frac{I}{d} + \frac{1}{d} \sum_{j=1}^{d^2-1} r_j \Gamma_j,\qquad r_j\in\mathbb{R},
\label{GBVdefinition}
\end{equation}
where \(I\) is the \(d\times d\) identity, and \(\Gamma_j\) are the generalized Pauli matrices constructed from tensor products of the standard Pauli matrices, satisfying \(\operatorname{Tr}(\Gamma_j)=0\) and \(\operatorname{Tr}(\Gamma_j\Gamma_k)=d\delta_{jk}\). The vector \(\vec{r}\coloneqq (r_1,\dots,r_{d^2-1})\) with \(r_j=\operatorname{Tr}(\rho\Gamma_j)\) is the GBV. 

The purity of \(\rho\) is \(\operatorname{Tr}(\rho^2)=\frac{1}{d}+\frac{1}{d}\sum_j r_j^2\), which depends only on the length of \(\vec{r}\). The normalized vector \(\hat{\mathbf{r}}=\vec{r}/|\vec{r}|\) defines the orientation of the state in GBV space. The tip of \(\hat{\mathbf{r}}\) lies on a \((d^2-2)\)-dimensional unit sphere—the generalization of the single-qubit Bloch sphere. The geometry of this sphere exhibits subtleties absent in the single‑qubit case; these are discussed in Appendix~\ref{appendix GBV} but are not essential for the present development.

The difference between two states \(\rho\) and \(\sigma\) thus decomposes into a purity difference and an orientational difference. A full description of the orientational difference would require as many variables as the dimension of the generalized Bloch sphere, which quickly becomes intractable even for moderate \(N\). To obtain a convenient and informative measure, we consider a single parameter that captures the overall orientational deviation:
\begin{equation}
\theta_{\mathrm{GBV}}(\rho, \sigma) = \arccos \left(\hat{\mathbf{r}}(\rho) \cdot \hat{\mathbf{r}}(\sigma) \right).
\label{GBVangle}
\end{equation}
This angle is defined when both GBVs are nonzero; the maximally mixed state has no defined orientation. It equals the geodesic distance between the two normalized GBVs on the generalized Bloch sphere, providing a coordinate‑independent measure of orientational misalignment.

Conventional metrics such as fidelity and trace distance quantify the overall dissimilarity between two states. They collapse all deviations into a single scalar, conflating orientational deviation with purity loss and lacking specificity for error diagnosis. The Fubini–Study and Bures distances ~\cite{bengtsson2017geometry} induce Riemannian geometries on the quantum state manifold and capture parameter-dependent structures such as quantum Fisher information, yet they are often too complex for practical error diagnosis. In contrast, the GBV angle isolates the orientational difference between states; combined with the GBV length that encodes purity, it directly reveals how different types of errors deform the state.

Beyond its physical intuition, the GBV representation is also practically useful because each component $r_j=\operatorname{Tr}(\rho\Gamma_j)$ is a directly measurable Pauli expectation value. When an algorithmic output is determined by a subset of such expectation values, as in the phase-estimation experiment considered here, full state tomography is unnecessary. Monitoring these task-relevant components directly reveals how errors perturb the quantities of interest. However, the number of required components may still grow substantially with system size for some tasks. The scalability of GBV diagnostics is therefore task dependent, and we do not claim that the framework provides a generally scalable diagnostic for arbitrary quantum circuits.

\begin{figure*}[htbp]
    \centering
    \includegraphics[scale=0.34]{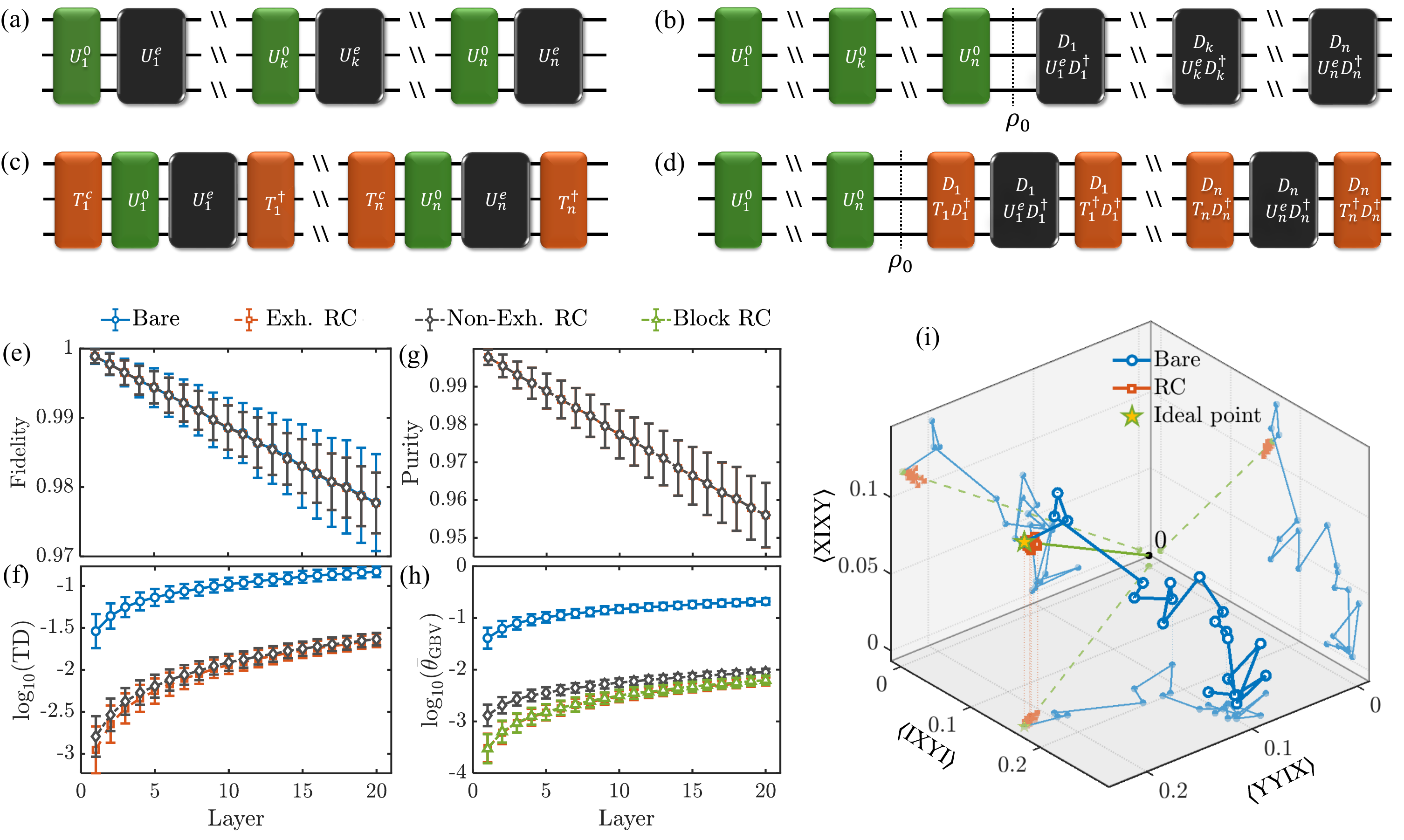}
   \caption{Numerical simulation of RC in the GBV representation. (a) A generic $n$-layer circuit with ideal evolutions $U_j^0$ and coherent errors $U_j^e$. (b) The equivalent effective-error-frame circuit (Appendix~\ref{appendix RC-channel}), where $D_j=\prod_{k=n}^{j+1}U_k^0$ and the ideal evolution first produces $\rho_0$. (c), (d) A representative RC sequence and its effective-error-frame form. (e)--(h) Fidelity, trace distance, purity, and $\theta_{\mathrm{GBV}}$ versus circuit depth for $N=4$, comparing bare, exhaustive-RC, non-exhaustive-RC, and block-RC circuits. (i) Projection of three representative GBV components onto a three-dimensional subspace; the origin is the maximally mixed state.}
    \label{fig:GBVA}
\end{figure*}

\subsection{Simulation results and analysis \label{Simulation}}

RC operates by inserting randomly chosen single-qubit Pauli gates at specific points in a quantum circuit, constrained such that the ideal logical operation remains unchanged; averaging over multiple randomized compilations converts coherent errors into stochastic Pauli errors~\cite{wallman2016noise}. While in practice RC often targets the larger coherent errors typical of two-qubit gates, its mechanism is general. We therefore employ the GBV representation to analyze RC at the level of individual circuit layers, without differentiating gate types.

We first consider circuits with only coherent errors; realistic noise is treated in Secs.~\ref{section GHZ experiment} and \ref{section braiding experiment}. Under this restriction, an $N$-qubit circuit of $n$ layers can be decomposed into an ideal error-free evolution followed by an effective error evolution (Figs.~\ref{fig:GBVA}(a)--(d); see Appendix~\ref{appendix RC-channel}). In the latter part, the raw coherent error of each layer is dressed by the ideal unitaries of subsequent layers, known as the toggling-frame transformation. Assuming the raw errors of different layers are independent and isotropically distributed, the toggling-frame conjugation will leave their statistical distribution unchanged. Therefore, we can directly generate the dressed errors as independent random operators (Appendix~\ref{Appendix Simulation details}).

We generate 1000 independent instances. Each instance contains a Haar-random pure target state $\rho_0$ and a random error configuration (see Appendix~\ref{Appendix Simulation details}). For each instance we obtain the final density matrix for three cases: the bare circuit, exhaustive RC, and non-exhaustive RC (details of the latter two are given in Appendix~\ref{Appendix Simulation details}). The explicit density-matrix expansions for the bare and RC cases are derived in Appendix~\ref{appendix RC-channel}.  

From these density matrices we compute four quantities: fidelity, trace distance, purity, and $\theta_{\mathrm{GBV}}$ defined in Eq.~(\ref{GBVangle}). All simulation results for the $N=4$ case are summarized in Figs.~\ref{fig:GBVA}(e)--(i). Figure~\ref{fig:GBVA}(e) shows that as the circuit depth increases, the average fidelity decreases monotonically, with almost no discernible difference between the bare and RC circuits. Figure~\ref{fig:GBVA}(f) plots the trace distance, where a clear separation emerges: RC suppresses the trace distance significantly. We refer to Appendix~\ref{Appendix Simulation details} for an explanation of why fidelity and trace distance behave differently.

Figures~\ref{fig:GBVA}(g) and (h) show the two GBV metrics: purity and $\theta_{\mathrm{GBV}}$. The bare circuit contains only coherent errors, so its purity remains exactly $1$ and is omitted. In RC circuits, purity decays nearly linearly with depth, reflecting the conversion of coherent errors into incoherent ones. The difference between exhaustive and non-exhaustive RC is barely visible. The behavior of $\theta_{\mathrm{GBV}}$ differs sharply between the bare and RC circuits: after 20 layers, the orientational deviation in the bare circuit is more than 30 times larger than in the RC circuits, while the purity of the RC state has dropped only about $5\%$. At the state level, $\theta_{\mathrm{GBV}}$ quantifies the orientational deviation of the output from the ideal state. Its small value under RC shows that the ensemble-averaged output exhibits an approximately isotropic contraction for the sampled input states and error configurations. The ratio $\theta_{\mathrm{GBV}}^{\mathrm{RC}}/\theta_{\mathrm{GBV}}^{\mathrm{bare}}$ directly quantifies the suppression of orientational deviation. At the channel level, Appendix~\ref{appendix channel vector} defines the input-independent angle $\theta_{\mathrm{ch}}$ and, for the parameters of Fig.~\ref{fig:GBVA}, gives $\theta_{\mathrm{ch}}(\mathcal T_{\mathrm{RC}})\le0.05168~\mathrm{rad}$ with probability at least $95\%$.

We further examine the dynamics of individual GBV components. Figure~\ref{fig:GBVA}(i) tracks three representative components with circuit depth. In the bare circuit, all three fluctuate with depth; under RC, they decay toward zero at similar rates, indicating a near-isotropic contraction at the state level. A single-layer Pauli twirl generally produces an anisotropic stochastic Pauli channel, so this behavior does not follow from Pauli twirling alone. Exhaustive RC eliminates the first-order coherent contributions and the inter-layer cross terms through second order, leaving only the layerwise second-order contributions (Appendix~\ref{appendix RC-channel}). Appendix~\ref{appendix convergence} gives the resulting composition of the Pauli-transfer eigenvalues, while Appendix~\ref{appendix channel vector} bounds their direction-dependent fluctuations under independent, isotropically distributed weak coherent errors. These results provide a channel-level explanation for the state-level near-isotropic contraction observed here. Finally, the simulations use $M=1000$ random sequences to
approximate the exhaustive-RC average; Appendix~\ref{appendix non-EX RC} shows that the corresponding relative root-mean-square sampling error is bounded by $M^{-1/2}\approx0.0316$.

\subsection{Block-RC \label{Block-RC}}

Standard layer-wise RC requires independently selecting and compiling twirls at every layer of each randomized circuit. To reduce this per-layer compilation and control overhead, we propose a block-RC protocol that correlates twirls across consecutive layers: all internal Pauli insertions cancel, leaving only pre-computable boundary operations.

\textit{Mechanism.} In the conventional layer-wise protocol, each layer $k$ carries an independent twirl $T_k$ and its compensator $T_k^c$. In block-RC, the twirls are correlated across layers: for $k>1$, $T_k$ is chosen as $T_{k-1}^{c\dagger}$, so that all internal Pauli insertions cancel. The sequence simplifies to a single effective twirl at the block entrance and a compensator at the exit, $T_n^c U_n U_{n-1}\cdots U_1 T_1$. For an exhaustive average over an $n$-layer, $N$-qubit block, this reduces the number of possible twirling configurations from $4^{nN}$ to $4^N$.

\textit{Applicability.} Because the Pauli group is normalized by the Clifford group, any Pauli operator conjugated through a Clifford block remains a Pauli operator; this ensures that the internal cancellation is exact for Clifford blocks. Block-RC therefore applies directly to blocks composed of Clifford gates. Non-Clifford gates need not be excluded from the circuit, but must define the boundaries between twirled blocks. The protocol is consequently most useful for Clifford-dominated circuits, including many stabilizer-based QEC routines, whereas frequent non-Clifford operations limit the achievable block size and hence the reduction in compilation overhead.

\textit{Limitation.} The resource savings come at a cost in error randomization. In layer-wise RC, every layer is individually twirled, so coherent errors are randomized at the finest granularity. In block-RC, errors accumulated inside the block are not randomized until the block boundary; a very large block therefore leaves more room for intra-block coherent accumulation. The optimal block size balances classical overhead against this residual orientational deviation. For the random-error model and circuit depths studied in Fig.~\ref{fig:GBVA}(h), even single-block twirling performs nearly indistinguishably from layer-wise RC; this observation is model dependent and does not imply that arbitrary block sizes are generally effective. Appendix~\ref{appendix Block-RC} examines the extreme case of identical coherent errors in every layer and shows how their accumulation within a single block degrades performance relative to layer-wise RC (Fig.~\ref{fig:block_rc_correlated_errors}).

\textit{Practical value.} By reducing per-layer randomized-circuit compilation and control operations, block-RC is most useful when these costs are a bottleneck, particularly in deep, repeatedly executed Clifford-dominated circuits. Section~\ref{statistics} compares block-RC with layer-wise RC under the noise models used in the topological-braiding simulations.

\begin{figure*}[htbp]
    \centering
    \includegraphics[scale=0.62]{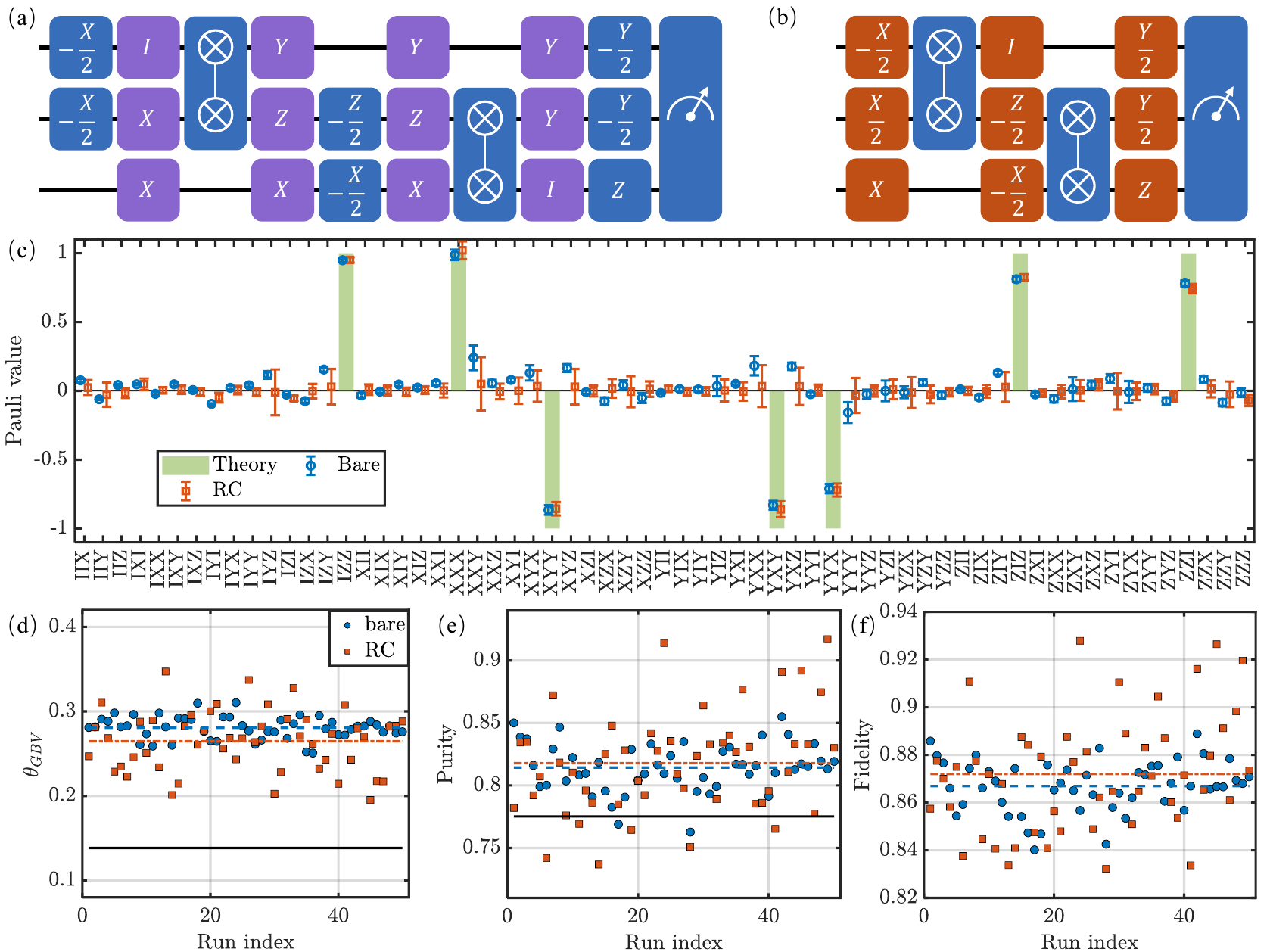}
   \caption{RC in GHZ state preparation. (a) A representative RC sequence. Blue blocks (single-qubit and two-qubit iSWAP gates) constitute the bare circuit; purple blocks are inserted twirling and compensation gates. (b) Recompiled circuit after absorbing the inserted gates into the single‑qubit gates of the bare circuit. (c) Expectation values of all 63 GBV components of a three-qubit GHZ state: ideal (light green bars), bare circuit (blue circles) and RC circuit (red squares). (d)–(f) $\theta_{\mathrm{GBV}}$, purity, and fidelity of individual sequences (RC) or repetitions (bare). Dashed lines indicate the mean of the individual values; solid lines indicate the values for the ensemble‑averaged RC state.}
    \label{fig:GHZ experiment}
\end{figure*}

\section{RC improves orientation under realistic noise \label{section GHZ experiment}}

In this section, we study how the geometric picture of RC established in Sec.~\ref{section GBV representation} changes under realistic noise, where decoherence is non-negligible. We address this through a three-qubit GHZ-state-preparation experiment on a superconducting quantum processor (see Appendix~\ref{appendix Exp-setup} for parameters; qubits Q2, Q3, and Q4 are used). 

In Fig.~\ref{fig:GHZ experiment}(a), the blue blocks form the bare circuit that transforms the initial state $\ket{000}$ to $\frac{1}{\sqrt{2}}(\ket{000} + \ket{111})$, followed by quantum state tomography (QST). We first use standard randomized benchmarking (RB) and purity-RB to quantify total gate infidelity and its incoherent fraction, respectively (Appendix~\ref{appendix GHZ analysis}). Single-qubit gates exhibit high fidelity ($>0.999$) with the contribution of incoherent errors to gate infidelity $>94\%$. The two-qubit iSWAP gate has fidelity $0.9917$, with about $84\%$ of the gate infidelity attributed to incoherent errors.

We first perform 50 independent experimental runs for the bare circuit, each with 3000 repetitions. For RC, 50 distinct twirling sequences are generated (one representative sequence is shown in Fig.~\ref{fig:GHZ experiment}(a)); each is executed 3000 times. This procedure produces an ensemble of 100 reconstructed density matrices. From each matrix, we extract the 63 GBV components; for each component, we then compute the mean and standard deviation across the 50 bare runs, and separately across the 50 RC sequences.

Figure~\ref{fig:GHZ experiment}(c) presents the GBV components from three sources: the experimental bare-circuit (blue circles with error bars) and RC (red squares with error bars) data obtained as described above, and the ideal three-qubit GHZ state (light green bars).

We first examine the two key diagnostic metrics—the GBV angle and purity—by computing them for each of the 50 bare runs and 50 RC sequences. Their distributions are shown in Figs.~\ref{fig:GHZ experiment}(d) and (e). The mean values of \(\theta_{\mathrm{GBV}}\) and purity (dashed lines) are nearly identical for the bare and RC circuits, consistent with RC redistributing the coherent-error contribution across sequences rather than systematically reducing the error of every individual sequence. RC nevertheless increases the variance of both metrics. For the bare circuit, variance arises from finite-sampling statistics under stable decoherence (residual fluctuations after averaging 3000 shots per run). Under RC, an additional sequence-to-sequence variation arises from the randomized coherent-error contribution. Thus, RC does not systematically improve either metric at the individual-sequence level.

The benefit of RC appears only in the ensemble‑averaged state. Averaging the 50 density matrices from RC yields a state whose \(\theta_{\mathrm{GBV}}\) is 0.138 (solid line in Fig.~\ref{fig:GHZ experiment}(d)), a 51\% reduction from the bare mean of 0.281. The purity (solid line in Fig.~\ref{fig:GHZ experiment}(e)) falls from 0.818 to 0.775, a reduction of 5\%. By contrast, the ensemble‑averaged state of the 50 bare runs has a \(\theta_{\mathrm{GBV}}\) of 0.261, a mere reduction of 7\%.  

Figure~\ref{fig:GHZ experiment}(f) shows the fidelity of individual runs. The mean fidelity of the bare circuit (86.7\%, blue dashed line) nearly matches that of RC (87.2\%, orange dashed line). For both bare and RC circuits, the fidelity of the ensemble-averaged state coincides with the mean fidelity of the individual runs. This observation illustrates a limitation of fidelity as a diagnostic tool. Because the target state $\rho_0$ is pure,
$F(\bar{\rho},\rho_0)=\frac{1}{50}\sum_{i=1}^{50}F(\rho_i,\rho_0)$, so fidelity does 
not distinguish whether errors of individual runs are systematic (bare) or randomized (RC). 

These results allow us to identify the dominant source of the orientational deviation in this device. RB and purity-RB show that the gate infidelity arises predominantly from incoherent errors (mainly dephasing). Under the dephasing-dominated noise model considered here, Pauli twirling leaves the Pauli-dephasing contribution intact while randomizing the coherent contribution~\cite{wallman2016noise}. The observed $51\%$ reduction in $\theta_{\mathrm{GBV}}$ therefore indicates that the small coherent fraction is responsible for the majority of the orientational deviation. This disproportionate impact is not resolved by fidelity alone but is directly exposed by the GBV diagnostics.

\section{RC Improves Accuracy of Phase Estimation \label{section braiding experiment}}

\begin{figure*}[htbp]
    \centering
    \includegraphics[scale=0.6]{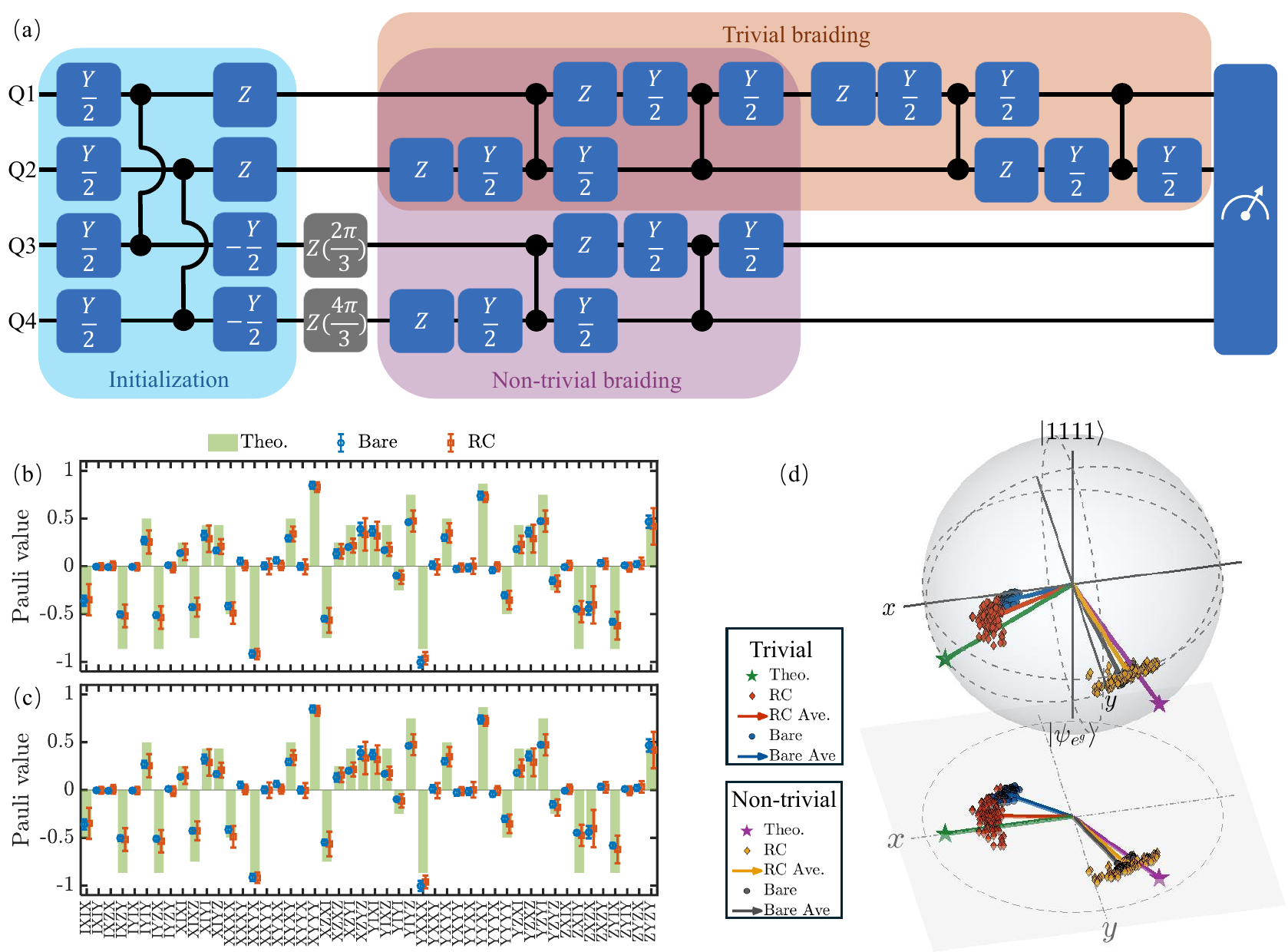}
    \caption{RC protects phase estimation in quantum simulation of topological braiding. (a) Bare circuit for the \(D(\mathbb{Z}_3)\) quantum double model (RC twirling gates omitted). All two‑qubit gates are Controlled‑Z (CZ). Total duration: 780~ns (trivial braiding) and 475~ns (nontrivial braiding). (b), (c) GBV components in the task‑relevant subspace for trivial and nontrivial braiding: ideal (light green bars), bare circuit (blue circles), RC (red squares). (d) Bloch sphere of the effective qubit spanned by \(\ket{\psi_{e^g}}\) and \(\ket{1111}\). Scattered symbols: final states of individual bare repetitions and individual RC sequences.}
    \label{fig:braiding1}
\end{figure*}

The preceding section demonstrates that RC improves the GBV orientation of quantum states even in the presence of significant decoherence. This improvement is particularly valuable for phase-sensitive algorithms, where the output is encoded in the orientation of a quantum state. In this section, we examine a phase-sensitive task: simulation of topological braiding in the \(D(\mathbb{Z}_3)\) quantum double model. This model hosts Abelian anyons whose braiding imparts topological phases that manifest as distinct GBV orientations of the final state, making it an ideal setting to demonstrate how RC's orientational protection translates into algorithmic advantage.

\subsection{Physical model and circuit implementation \label{DZ3 model}}

A minimal version of the \(D(\mathbb{Z}_3)\) model can be simulated with two qutrits (Appendix~\ref{appendix braiding theory}). In our experiment, each qutrit is encoded in two superconducting qubits. The basis states \(\ket{00}\), \(\ket{01}\), and \(\ket{10}\) of the two qubits are mapped to the qutrit states \(\ket{\widetilde{0}}\), \(\ket{\widetilde{1}}\), and \(\ket{\widetilde{2}}\), respectively. As shown below, the braiding operation imparts a global phase to the encoded two-qutrit subspace, which is unobservable within the subspace. However, in the larger Hilbert space, this phase becomes a relative one between the subspace and a reference state outside it, provided the two parts acquire different phases during braiding. We use the four-qubit product state \(\ket{1111}\) as the reference, which is unaffected by braiding.

Figure~\ref{fig:braiding1}(a) shows the experimental circuit. Qubits Q1 and Q2 encode Qutrit 1; Q3 and Q4 encode Qutrit 2. The initialization prepares a superposition of the ground state of the two‑qutrit \(D(\mathbb{Z}_3)\) model \(\ket{\psi_g} = \frac{1}{\sqrt{3}}(\ket{\widetilde{00}}+\ket{\widetilde{11}}+\ket{\widetilde{22}})\) and the reference state \(\ket{1111}\):
\begin{align}
\ket{\psi_{\mathrm{ini}}}&=\frac{\sqrt{3}}{2}\ket{\psi_g}+\frac{1}{2}\ket{1111}\notag\\
&=\frac12\ket{0000}+\frac12\ket{0101}+\frac12\ket{1010}+\frac12\ket{1111}.
\label{eq:initial state}
\end{align}
Two virtual \(Z\) gates with phases \(\frac{2\pi}{3}\) and \(\frac{4\pi}{3}\) are then applied to qubits Q3 and Q4, respectively, simulating the creation of an electric anyon on Qutrit 2. These operations are confined to the encoded two‑qutrit subspace and leave the reference state \(\ket{1111}\) unchanged. The subspace is thus prepared into the eigenstate \(\ket{\psi_{e^g}}\) of the \(D(\mathbb{Z}_3)\) model, and the full system state becomes
\begin{align}
\ket{\psi_1}&=\frac{\sqrt{3}}{2}\ket{\psi_{e^g}}+\frac{1}{2}\ket{1111}\notag\\
&=\frac12\ket{0000}+\frac12 e^{i\frac{2\pi}{3}}\ket{0101}+\frac12 e^{i\frac{4\pi}{3}}\ket{1010}+\frac12\ket{1111}.
\label{eq:eg state}
\end{align}

The trivial braiding circuit (orange shaded block in Fig.~\ref{fig:braiding1}(a)) simulates the creation and annihilation of a magnetic anyon without braiding with the electric anyon mentioned above, yielding a phase of 0. The final state therefore remains unchanged, as given in Eq.~(\ref{eq:eg state}). On the other hand, the nontrivial braiding circuit simulates a half-braiding between the magnetic and electric anyons \cite{Uniquely2023Hai}, leading to a phase of $\frac{2\pi}{3}$. After this operation, the full system state is
\begin{align}
\ket{\psi_2}
&= \frac{\sqrt{3}}{2} e^{i\frac{2\pi}{3}} \ket{\psi_{e^g}} + \frac{1}{2} \ket{1111} \notag \\
&= \frac{1}{2} e^{i\frac{2\pi}{3}}\ket{0000}
 + \frac{1}{2} e^{i\frac{4\pi}{3}} \ket{0101}
 + \frac{1}{2} \ket{1010}
 + \frac{1}{2} \ket{1111}.
 \label{eq:nontrivial state}
\end{align}

The braiding phase can be extracted from the density matrix $\rho$ reconstructed by QST using
\be
\phi = -\arg\bigl[\operatorname{Tr}(\rho \ket{\psi_{e^g}}\bra{1111})\bigr]. \label{eq:phi}
\ee
\noindent However, QST is not necessary. The operator $\ket{\psi_{e^g}}\bra{1111}$ contains only 48 non‑zero components when expanded in the generalized Pauli basis: $\ket{\psi_{e^g}}\bra{1111} = \sum_{j=1}^{48} p_j \Gamma_j$ (the corresponding Pauli operators are listed in Figs.~\ref{fig:braiding1}(b) and (c) and Table~\ref{tab:pauli_expansion} in Appendix~\ref{appendix braiding theory}). These 48 operators span the task‑relevant subspace, within which both ideal final states have 32 non‑zero components, indicated by light green bars in Figs.~\ref{fig:braiding1}(b) and (c).

Measuring the components in this subspace suffices to determine $\phi$. Indeed, Eq.~(\ref{eq:phi}) simplifies to
\be
\phi = \operatorname{atan2}\left(-\sum_{j=1}^{48}\operatorname{Im}(p_j)r_j^{\rho},\sum_{j=1}^{48}\operatorname{Re}(p_j)r_j^{\rho}\right), \label{eq:phi2}
\ee
where $r_j^{\rho} = \operatorname{Tr}(\rho \Gamma_j)$ is the $j$-th GBV component. This expression suggests that any technique improving the alignment of the GBV with the ideal orientation may enhance the accuracy of $\phi$, since $\phi$ is sensitive to the relative weights of these components.

\subsection{Experimental data analysis \label{braiding data analysis}}

The experiments were performed on the same superconducting quantum processor used in Sec.~\ref{section GHZ experiment}. Four physically connected qubits forming a closed loop implement two qutrits (Fig.~\ref{fig:braiding1}(a); also see Appendix~\ref{appendix Exp-setup}). We execute the circuit over 100 independent random RC sequences; the bare circuit is repeated 100 times for comparison. Figures~\ref{fig:braiding1}(b) and (c) show the measured 48 components for trivial and nontrivial braiding. 

For each data set, we first average all 100 density matrices and then evaluate the phase using Eq.~(\ref{eq:phi2}). Statistical uncertainties are estimated by delete-one jackknife resampling: each sample omits one complete RC sequence or bare-circuit repetition, and the phase is recalculated from the remaining average state. Appendix~\ref{appendix jackknife analysis} gives the estimator. All phase uncertainties reported below are one standard error (SE).

\begin{table}[ht]
\centering
\caption{Metrics extracted from experimental results. Phases are evaluated from the average density matrices and are reported in radians with one jackknife standard error.}
\label{tab:braiding_results}
\begin{tabular}{c c c c c c}
\hline\hline
Braiding Type & Circuit & Fidelity & Purity & $\theta^{\mathrm{48-D}}_{\mathrm{GBV}}$ & $\phi$\\
\hline
\multirow{2}{*}{Trivial}
 & Bare & 0.598 & 0.500 & 0.524 & $-0.690\pm0.011$\\
 & RC & 0.664 & 0.468 & 0.213 & $-0.227\pm0.025$\\
\hline
\multirow{2}{*}{Nontrivial}
 & Bare & 0.771 & 0.658 & 0.280 & $1.9197\pm0.0057$\\
 & RC & 0.772 & 0.635 & 0.245 & $1.988\pm0.021$\\
\hline\hline
\end{tabular}
\end{table}

Table~\ref{tab:braiding_results} summarizes the relevant metrics. For the trivial braiding circuit, RC reduces the 48-dimensional GBV angle $\theta^{\mathrm{48-D}}_{\mathrm{GBV}}$ from 0.524 to 0.213, a substantial improvement that directly translates into a significantly more accurate phase estimate. In the nontrivial case, the improvement in $\theta^{\mathrm{48-D}}_{\mathrm{GBV}}$ is modest---from 0.280 to 0.245---and correspondingly, the phase improvement is also less pronounced.

Assuming independent bare and RC data sets, the reductions in absolute phase error correspond to approximately $16.8$ and $3.1$ combined SEs for trivial and nontrivial braiding, respectively. Thus, the phase improvement is statistically supported in both cases despite larger uncertainties under RC.

\begin{table*}[htbp]
\caption{Fractions of 1000 random coherent-error configurations satisfying each condition. Layer-wise RC and block-RC are applied to both the trivial and nontrivial circuits.}
\centering
\begin{tabular}{lcccc}
\hline
Condition & 
Layer-wise RC (Trivial) & 
Layer-wise RC (Nontrivial) & 
Block-RC (Trivial) & 
Block-RC (Nontrivial) \\
\hline
$\mathrm{Fidelity}_{\mathrm{bare}} < \mathrm{Fidelity}_{\mathrm{RC}}$ 
& 0.445 & 0.449 & 0.462 & 0.457 \\

$\mathrm{Purity}_{\mathrm{bare}} > \mathrm{Purity}_{\mathrm{RC}}$ 
& 1.000 & 1.000 & 1.000 & 1.000 \\

$\theta^{\mathrm{48-D}}_{\mathrm{GBV}}(\mathrm{bare}) 
> \theta^{\mathrm{48-D}}_{\mathrm{GBV}}(\mathrm{RC})$ 
& 1.000 & 1.000 & 1.000 & 1.000 \\

$|\Delta\phi|_{\mathrm{bare}} > |\Delta\phi|_{\mathrm{RC}}$ 
& 0.928 & 0.938 & 0.936 & 0.931 \\

\hline
\end{tabular}
\label{tab:RC_probability_comparison}
\end{table*}

Next, we project the high-dimensional state onto the effective two-dimensional subspace spanned by $\ket{\psi_{e^g}}$ and $\ket{1111}$, yielding an effective qubit whose Bloch sphere representation is shown in Fig.~\ref{fig:braiding1}(d). Although RC increases the spread of individual points, its average is significantly closer in phase to the ideal than the bare average. The improved phase accuracy follows directly from the reduced systematic offsets in the 48 components.

\subsection{Statistics across different error configurations \label{statistics}}

The above experimental results represent only a single error configuration, as both the decoherence and the specific coherent errors are fixed by the device and the particular circuit instance. Therefore we employ numerical simulations to investigate whether RC remains effective under different error configurations.

Following the same procedure in Sec.~\ref{section GBV representation}, we generate 1000 random coherent error configurations, assuming that only two-qubit gates are subject to coherent errors while single-qubit gates are ideal---an assumption consistent with our independently benchmarked gate fidelities. The decoherence channels are fixed to those experimentally characterized for our device. For each configuration, we simulate both the bare (100 repetitions) and the RC (100 random sequences) circuits. From the resulting 1000 pairs of final states, we compute the same set of metrics as in the experiment. The statistical distributions are summarized in Table~\ref{tab:RC_probability_comparison}.

Within all 1000 sampled configurations, RC reduces purity and improves $\theta^{\mathrm{48-D}}_{\mathrm{GBV}}$, showing that its orientational protection is robust across the error ensemble studied here. By contrast, the fraction of configurations in which RC improves fidelity is close to 0.5, indicating that fidelity is largely insensitive to the structural changes induced by RC.

For algorithmic performance, RC reduces the phase error in over 92\% of configurations. The small fraction of cases without improvement ($\sim$7\%) may correspond to cases where orientation improvement alone is insufficient, or signal degradation is particularly severe. Both trivial and nontrivial cases show similar statistics, with differences well within statistical fluctuations. 

The simulation also shows that conventional layer-wise RC and block-RC (Sec.~\ref{Block-RC}) yield nearly identical statistics for all metrics under the noise models considered here. Thus block-RC can reproduce the observed error suppression while reducing per-layer compilation overhead in circuit structures satisfying the conditions described in Sec.~\ref{Block-RC}.

\subsection{Dephasing wall: a limit of RC \label{Dephasing wall}}

\begin{figure}[htbp]
    \centering
    \includegraphics[scale=0.45]{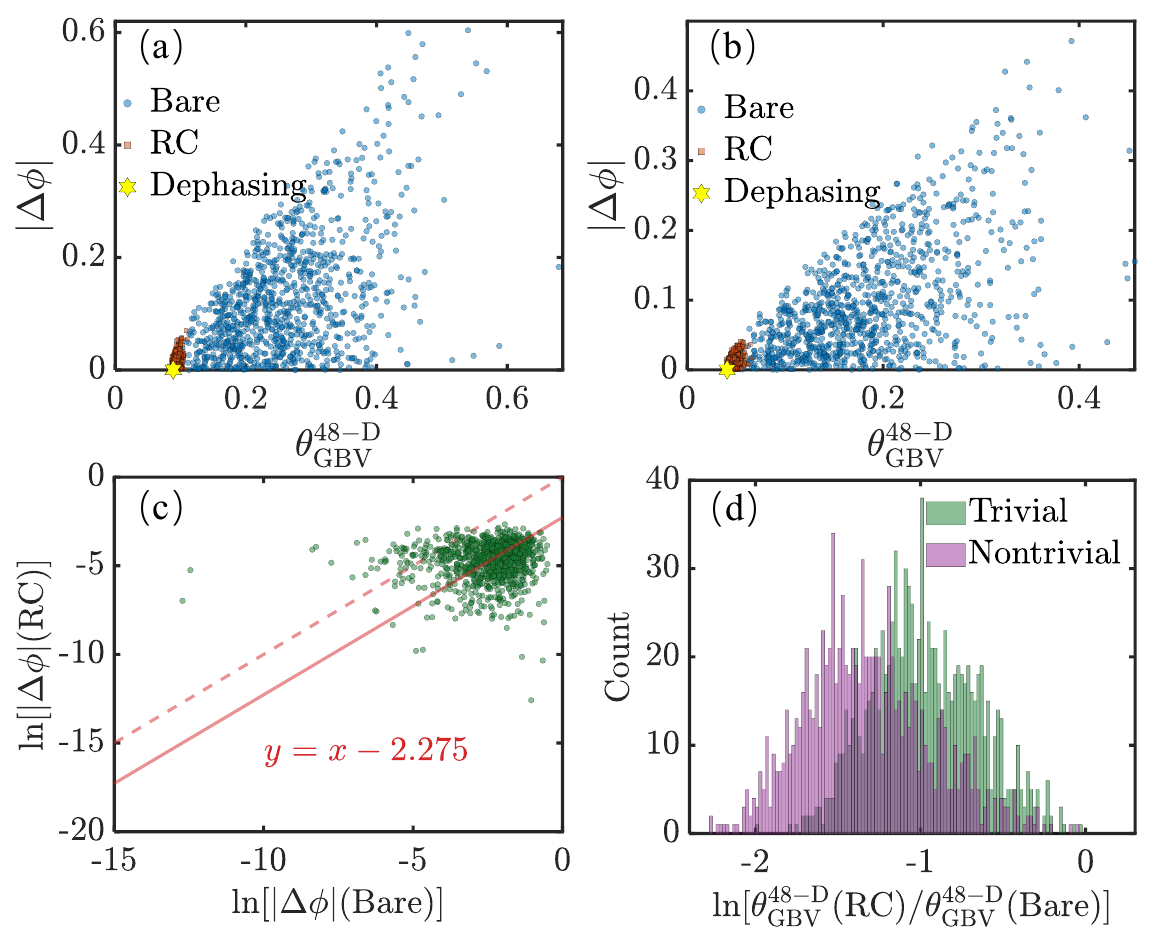}
    \caption{Numerical simulation of the topological braiding circuit across 1000 error configurations. (a), (b) Phase error $|\Delta\phi|$ versus $\theta^{\mathrm{48-D}}_{\mathrm{GBV}}$ for trivial and nontrivial braiding; yellow hexagrams denote dephasing-only simulations. (c) RC versus bare phase errors for trivial braiding. In the natural-logarithmic coordinates shown, the fitted intercept is $-2.275$, corresponding to a reduction factor of $e^{2.275}\approx9.7$ inferred from the fit. (d) Histogram of $\ln[\theta^{\mathrm{48-D}}_{\mathrm{GBV}}(\mathrm{RC})/\theta^{\mathrm{48-D}}_{\mathrm{GBV}}(\mathrm{bare})]$.}
    \label{fig:braiding2}
\end{figure}

To further understand these effects, we analyze the layer-wise RC simulation data. Figures~\ref{fig:braiding2}(a) and (b) provide a global view of RC's impact for trivial and nontrivial cases, respectively. Each panel contains 1000 pairs of data points, each pair consisting of the bare and RC results for the same error configuration. In each panel, a yellow hexagram marks a reference simulation with dephasing only, using the experimentally calibrated dephasing time.

For the trivial case, the RC points are tightly clustered around $\theta\approx0.09$ and $|\Delta\phi|\approx0.02$. In contrast, the bare points span a much larger range. The nontrivial case shows even tighter clustering at $\theta\approx0.04$, again in stark contrast to the bare distribution.

The bare points are scattered because each error configuration pushes the state in a different orientation; their diversity reflects the variety of coherent errors. RC effectively collapses these disparate trajectories, erasing most of the configuration-to-configuration variation. The residual offset of each RC cluster represents the orientational error that persists after correction. It has two origins: dephasing, which is unchanged by Pauli twirling in our noise model, and residual coherent errors that survive RC. However, the simulations in Sec.~\ref{section GBV representation} (see Fig.~\ref{fig:GBVA}(h)) show that RC reduces the GBV angle well below $0.01$ in the absence of decoherence, much smaller than the offset observed here. Hence the offset is dominated by dephasing.

\begin{figure*}[!t]
    \centering
    \includegraphics[scale=0.3]{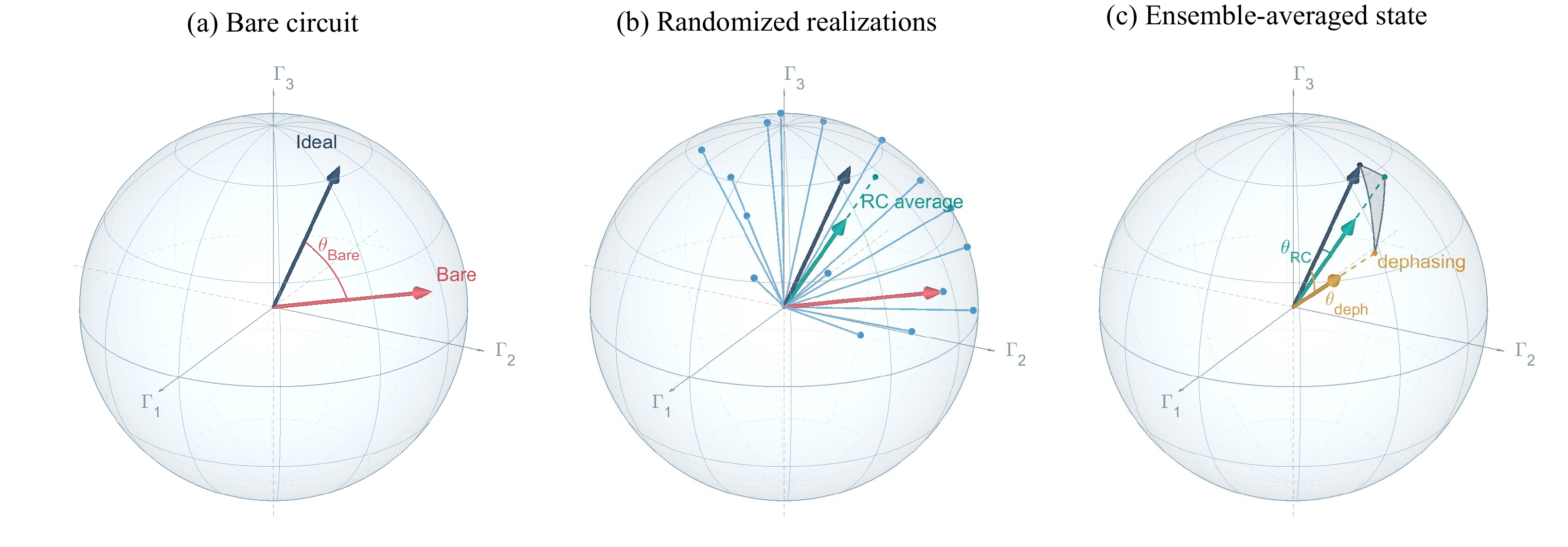}
    \caption{Schematic GBV picture of RC, shown in a three-dimensional projection. Arrow direction and length represent orientation and purity, respectively. (a) Coherent error rotates the bare GBV from the ideal direction by $\theta_{\mathrm{Bare}}$ without changing its length. (b) RC randomizes the individual GBV orientations (blue); the bare result (red) is the identity-twirling member, while their average (green) is better aligned with the ideal direction and slightly shortened. (c) RC thus suppresses orientational deviation with modest purity loss, whereas dephasing causes additional contraction and a residual deviation that RC cannot remove.}
    \label{fig:RC-summary}
\end{figure*}

The longer trivial circuit (780 ns) exhibits a larger residual than the shorter nontrivial circuit (475 ns), consistent with greater exposure to dephasing, although the two circuits also differ in structure. Together with the dephasing-only reference points, this observation identifies a dephasing-dominated residual—a limit beyond which RC alone cannot further improve orientation under the noise model considered here. Appendix~\ref{appendix dephasing wall} gives a perturbative account of this dephasing wall. In the absence of decoherence, the corresponding state-level behavior reduces to the near-isotropic contraction found in Sec.~\ref{section GBV representation}, whose channel-level deviation from the depolarizing direction is quantified in Appendix~\ref{appendix channel vector}.

We also compute $\theta^{\mathrm{48-D}}_{\mathrm{GBV}}$ between the pure-dephasing and RC-averaged states. Over the 1000 configurations, this angle is $0.0259 \pm 0.0096$ (trivial) and $0.019 \pm 0.0081$ (nontrivial), showing that the two states are closely, but not exactly, aligned. This residual misalignment accounts for the small finite $\Delta\phi$ of the RC-averaged states, while their lower purity due to ensemble mixing may further degrade the signal-to-noise ratio under finite sampling.

Figure~\ref{fig:braiding2}(c) compares $|\Delta\phi|$ before and after RC for the trivial braiding circuit (the nontrivial case shows similar behavior). Each point corresponds to one error configuration. The majority of points lie well below the diagonal $y=x$ (dashed line), showing that RC improves the phase estimate. The average phase error over 1000 configurations drops from $0.130$ (bare) to $0.013$ (RC).

In approximately 7\% of the configurations, RC slightly increases the phase error (points above the dashed line). All such points, however, are confined to a region where both bare and RC errors are very small ($|\Delta\phi|_{\mathrm{bare}} < 0.04, |\Delta\phi|_{\mathrm{RC}} < 0.06$). This behavior arises from specific alignments between the error structure and the projection direction of the measurements. 

Figure~\ref{fig:braiding2}(d) shows a histogram of $\ln(\theta^{\mathrm{48-D}}_{\mathrm{GBV}}(\mathrm{RC}) / \theta^{\mathrm{48-D}}_{\mathrm{GBV}}(\mathrm{bare}))$ over the 1000 configurations. The distributions peak well below zero, indicating that RC efficiently reduces the orientational deviation relative to the bare circuit. On average, RC suppresses the deviation to about $39.69\%$ (trivial) and $28.75\%$ (nontrivial) of its bare-circuit value. Thus RC's orientational protection remains effective in the presence of the dephasing-dominated residual.

\section{Conclusion \label{section discussion outlook}}

Using the GBV representation, we have identified a cumulative effect of RC that does not follow from single-layer Pauli randomization alone. Under the independent, isotropically distributed weak coherent errors considered here, the accumulated ensemble-averaged RC channel lies close to the depolarizing direction and produces a state-level near-isotropic contraction, strongly suppressing orientational deviation with only modest purity loss. When dephasing is present, RC suppresses the coherent contribution but leaves a dephasing-dominated residual under the noise model considered here, giving rise to the dephasing wall. Figure~\ref{fig:RC-summary} summarizes this mechanism geometrically: coherent errors produce systematic orientational deviation, whereas averaging randomized realizations restores the GBV toward the ideal orientation at the cost of radial contraction.

The GHZ-state experiment verifies this purity--orientation trade-off on hardware, while the topological phase-estimation experiment demonstrates its algorithmic consequence: improved GBV orientation can yield more accurate phase estimates even when fidelity shows little improvement. RC should therefore be evaluated using quantities relevant to the target task rather than fidelity alone.

\section{Acknowledgement}
This work was supported by the National Natural Science Foundation of China (No. 12074166), the Guangdong Provincial Key Laboratory (Grant No. 2019B121203002), the Innovation Program for Quantum Science and Technology (2021ZD0301703), Guangdong Major Project of Basic Research (2025B0303000007), and the Shenzhen Science and Technology Program (Grant No. KQTD20200820113010023).

\clearpage

\appendix

\section{Geometric Structure of the Generalized Bloch Representation\label{appendix GBV}}

This appendix elaborates on the geometric structure of the set of physical states within the GBV space.

For a single qubit ($N=1$), the GBV representation reduces to the standard Bloch picture. The set of all physical density matrices corresponds exactly to the Bloch ball in $\mathbb{R}^3$: the pure states form the surface (the Bloch sphere, $\mathbb{S}^2$), and mixed states fill the interior. 

For $N \ge 2$ qubits, the situation is fundamentally different. The space of all density matrices is embedded in a real vector space of dimension $4^N-1$. The normalized GBV, $\hat{\mathbf{r}} = \vec{r}/\|\vec{r}\|$, lies on a unit sphere $\mathbb{S}^{4^N-2}$. However, the set of physical density matrices---those satisfying $\rho \ge 0$ and $\mathrm{Tr}(\rho)=1$---forms a convex subset of the ball enclosed by $\mathbb{S}^{4^N-2}$, and the set of pure states ($\rho^2 = \rho$) is a zero-measure subset of the sphere. Moreover, the pure-state submanifold is not dense on $\mathbb{S}^{4^N-2}$. There exist open regions on the sphere that contain no pure state. 

Despite such subtlety, the GBV representation remains both conceptually clear and practically useful. $\mathbb{S}^{4^N-2}$ provides a geometric stage on which every physical state—pure or mixed—is represented by a radial vector. Its length encodes purity, and, for a nonzero GBV, its orientation $\hat{\mathbf{r}}$ encodes the informational content of the state. Any two states with nonzero GBVs define a unique angle $\theta_{\mathrm{GBV}}$ via Eq.~(\ref{GBVangle}) of the main text, which measures their orientational separation on the sphere. Noise processes can then be visualized as movements of GBVs on or inside this sphere. The separation of orientation and purity is unaffected by the intricate geometry discussed above, and still enables the analysis of techniques like RC.

\section{Quantum Channel Analysis for RC\label{appendix RC-channel}}

In this appendix, we derive the density matrices for the bare and exhaustive RC circuits. The results explicitly show how RC eliminates coherent rotations to first order and removes inter‑layer cross terms to second order, thereby providing the algebraic foundation for the discussion in the main text.

Consider an $N$-qubit circuit of $n$ layers. The ideal unitary evolution and the coherent error of the $j$-th layer are denoted as $U_j^0$ and $U_j^e$, respectively. The overall evolution of the circuit can be written as $U_{\mathrm{bare}}=U_n^eU_n^0U_{n-1}^eU_{n-1}^0 \cdots U_1^eU_1^0$ and the ideal error-free evolution is $U_0=U_n^0U_{n-1}^0 \cdots U_1^0$.

Notice that $U_{\mathrm{bare}}$ can be rearranged as:
\[U_{\mathrm{bare}}=\prod_{j=n}^1 U_j^e U_j^0=(\prod_{j=n}^1 ((\prod_{k=n}^{j+1} U_k^0) U_j^e (\prod_{k=n}^{j+1} U_k^0)^{\dagger}))U_0.\]
\noindent Here the products are time ordered, with higher-index factors placed to the left. This formulation separates the evolutions induced by the coherent errors from the ideal evolution $U_0$ by moving all $U_j^e$ terms to the left (i.e., to a later chronological position), where they are conjugated by the ideal evolutions of all subsequent layers. Figure~\ref{fig:GBVA}(b) depicts this rearrangement. Therefore, to investigate the effect of the coherent errors, we only need to pay attention to the part $U_{e-\mathrm{bare}}=\prod_{j=n}^1 ((\prod_{k=n}^{j+1} U_k^0) U_j^e (\prod_{k=n}^{j+1} U_k^0)^{\dagger})$. The actual output state is:
\begin{align*}
    \rho_{\mathrm{bare}}&=U_{\mathrm{bare}}\rho_{\mathrm{in}} U_{\mathrm{bare}}^{\dagger} \notag=U_{e-\mathrm{bare}} U_0 \rho_{\mathrm{in}} U_0^{\dagger} U_{e-\mathrm{bare}}^{\dagger} \\
    &= U_{e-\mathrm{bare}} \rho_0 U_{e-\mathrm{bare}}^{\dagger}.
\end{align*}
Here $\rho_{\mathrm{in}}$ and $\rho_0$ stand for the input and ideal error-free output states, respectively.

The coherent error of the $j$-th layer is modeled as $U_j^e = \exp(-i\theta_j \boldsymbol{\Gamma}^{(j)})$. Here $\boldsymbol{\Gamma}^{(j)}$ denotes a $2^N \times 2^N$ traceless Hermitian matrix. For the bare circuit, the erroneous evolution operator $U_{e-\mathrm{bare}}$ thus becomes $U_{e-\mathrm{bare}} = \prod_{j=n}^1 \big( (\prod_{k=n}^{j+1} U_k^0) e^{-i\theta_j \boldsymbol{\Gamma}^{(j)}} (\prod_{k=n}^{j+1} U_k^0)^\dagger \big)$. We assume that the coherent errors in different layers are independent. Since the toggling-frame transformation $\prod_{k=n}^{j+1} U_k^0$ merely conjugates $\boldsymbol{\Gamma}^{(j)}$ unitarily (preserving its traceless Hermitian nature), its effect can be absorbed into the definition of $\boldsymbol{\Gamma}^{(j)}$. Thus, without loss of generality, we may write $U_{e-\mathrm{bare}} = \prod_{j=n}^1 e^{-i\theta_j \boldsymbol{\Gamma}^{(j)}}$, and the output density matrix of the bare circuit becomes
\[
\rho_{\mathrm{bare}} = \Bigl(\prod_{j=n}^1 e^{-i\theta_j \boldsymbol{\Gamma}^{(j)}}\Bigr) \rho_0 \Bigl(\prod_{j=1}^n e^{i\theta_j \boldsymbol{\Gamma}^{(j)}}\Bigr).
\]

Assume $\theta_j \ll 1$ and let $\theta=\max_j|\theta_j|$. Expanding $e^{-i\theta_j \boldsymbol{\Gamma}^{(j)}}$ to second order gives $e^{-i\theta_j \boldsymbol{\Gamma}^{(j)}} = I - i\theta_j \boldsymbol{\Gamma}^{(j)} - \frac{\theta_j^2}{2} \bigl(\boldsymbol{\Gamma}^{(j)}\bigr)^2 + \mathcal{O}(\theta_j^3)$. Substituting the expression into $\rho_{\mathrm{bare}}$ yields:
\begin{align}
\rho_{\mathrm{bare}} = &\rho_0 - i\sum_{j=1}^n \theta_j \bigl[\boldsymbol{\Gamma}^{(j)}, \rho_0\bigr] \notag\\
&-\frac{1}{2}\sum_{j=1}^n \theta_j^2 \Bigl( \bigl(\boldsymbol{\Gamma}^{(j)}\bigr)^2 \rho_0 + \rho_0 \bigl(\boldsymbol{\Gamma}^{(j)}\bigr)^2 - 2 \boldsymbol{\Gamma}^{(j)} \rho_0 \boldsymbol{\Gamma}^{(j)} \Bigr) \notag\\
&-\sum_{1\le j<k\le n} \theta_j\theta_k \Bigl( \boldsymbol{\Gamma}^{(k)}\boldsymbol{\Gamma}^{(j)} \rho_0 + \rho_0 \boldsymbol{\Gamma}^{(j)}\boldsymbol{\Gamma}^{(k)} \notag\\
&-\boldsymbol{\Gamma}^{(j)} \rho_0 \boldsymbol{\Gamma}^{(k)} - \boldsymbol{\Gamma}^{(k)} \rho_0 \boldsymbol{\Gamma}^{(j)} \Bigr) + \mathcal{O}(\theta^3). \label{baredensitymatrix}
\end{align}
\noindent The first line contains the first‑order terms, which describe coherent rotations of the state. The second line collects the second‑order contributions from each layer individually, while the last summation comprises coherent cross terms between different layers; their operator ordering reflects the generally noncommuting error generators.

For the RC case, we first consider the implementation using all possible sequences, termed ``exhaustive RC.'' In such an implementation, we apply all \(4^N\) possible Pauli twirling operations to each layer. Experimentally, this would require enumerating all \(4^{nN}\) sequences for $n$ layers. In simulation, however, we have access to the intermediate density matrices, allowing us to implement an equivalent layer‑wise twirling: apply all Pauli gates to the first layer, average over the resulting density matrices, then proceed to the next layer, and so on. 

We now show that exhaustive RC removes the first‑order contribution entirely. For a single layer, one applies a complete Pauli twirl: 
\[
\mathcal{T}(\rho) = \frac{1}{|\mathcal{P}|}\sum_{P\in\mathcal{P}} P^\dagger U^e P \,\rho\, P^\dagger U^{e\dagger} P.
\]
\noindent Here $\mathcal{P}$ is the $N$-qubit Pauli group (ignoring global phases). Expand $U^e$ to first order, $P^\dagger U^e P = I - i\theta P^\dagger \boldsymbol{\Gamma} P + \mathcal{O}(\theta^2)$. Substitute into $\mathcal{T}(\rho)$:
\[
\mathcal{T}(\rho) = \rho-i\theta\frac{1}{|\mathcal{P}|}\bigl[\sum_{P} P^\dagger \boldsymbol{\Gamma} P,\;\rho\bigr] + \mathcal{O}(\theta^2).
\]
\noindent The Pauli group is a unitary 1‑design, so for any traceless $X$, $\sum_{P} P^\dagger X P = 0$. Since $\boldsymbol{\Gamma}$ is traceless, the first‑order term vanishes:
\[
\mathcal{T}(\rho) = \rho + \mathcal{O}(\theta^2).
\]
\noindent Thus, exhaustive RC eliminates the first‑order contribution of any coherent error in a single layer.

To retain the second-order contribution, define $\boldsymbol{\Gamma}_P=P^\dagger\boldsymbol{\Gamma}P$. Expanding the same twirl to second order gives
\[
\mathcal{T}(\rho)=\rho-\frac{\theta^2}{2|\mathcal{P}|}\sum_{P\in\mathcal{P}}\Bigl(\boldsymbol{\Gamma}_P^2\rho+\rho\boldsymbol{\Gamma}_P^2-2\boldsymbol{\Gamma}_P\rho\boldsymbol{\Gamma}_P\Bigr)+\mathcal{O}(\theta^3).
\]
For the $j$-th layer, we use the notation $\boldsymbol{\Gamma}_P^{(j)}=P^\dagger\boldsymbol{\Gamma}^{(j)}P$.

For a multi‑layer circuit, each layer undergoes an independent Pauli twirl. The overall evolution is the composition of these twirled channels:
\[
\rho_{\mathrm{RC}} = \mathcal{T}_n \circ \mathcal{T}_{n-1} \circ \cdots \circ \mathcal{T}_1 (\rho_0),
\]
where $\mathcal{T}_j$ denotes the twirling operation on the $j$-th layer. Because each $\mathcal{T}_j$ eliminates this layer's error to the first order and the twirling operations are independent and act sequentially, the final density matrix contains no first‑order terms from any layer. Moreover, when expanding to second order, the cross terms between different layers are absent in the RC case, since such terms would require simultaneous first‑order contributions from two different layers. The only remaining second‑order contributions are those arising from each layer independently. In the following, we prove the above conclusions by induction. 

The induction hypothesis is: after applying exhaustive RC to the first $m$ layers, the resulting state $\rho_m$ contains no cross terms \(\theta_j\theta_k\) (\(j\neq k\)) up to second order. Moreover, it contains no first-order terms in any \(\theta_j\). Hence it can be written as
\[
\rho_m = \rho_0 + \sum_{j=1}^{m} \theta_j^2 A_j + \mathcal{O}(\theta^3),
\]
where each \(A_j\) is an operator that depends only on \(\boldsymbol{\Gamma}^{(j)}\) (and possibly on \(\rho_0\)), and the \(\mathcal{O}(\theta^3)\) term includes only cubic and higher powers of the small parameters.

For $m=1$, the conclusion follows from the single-layer result. Applying the $(m+1)$-th layer twirling \(\mathcal{T}_{m+1}\) to \(\rho_m\) gives
\begin{align*}
\mathcal{T}_{m+1}(\rho_m) &= \rho_m - \frac{\theta_{m+1}^2}{2|\mathcal{P}|}\sum_{P\in\mathcal{P}}\Bigl((\boldsymbol{\Gamma}^{(m+1)}_P)^2\rho_m + \rho_m(\boldsymbol{\Gamma}^{(m+1)}_P)^2 \\
&-2\boldsymbol{\Gamma}^{(m+1)}_P\rho_m\boldsymbol{\Gamma}^{(m+1)}_P\Bigr) + \mathcal{O}(\theta_{m+1}^3).
\end{align*}
Expanding to second order in all small parameters, we obtain
\[
\begin{aligned}
\rho_{m+1} &= \mathcal{T}_{m+1}(\rho_m) =
\Bigl[\rho_0 + \sum_{j=1}^{m}\theta_j^2 A_j\Bigr] 
- \frac{\theta_{m+1}^2}{2|\mathcal{P}|}\sum_{P\in\mathcal{P}} \\
&\Bigl((\boldsymbol{\Gamma}^{(m+1)}_P)^2\rho_0 + \rho_0(\boldsymbol{\Gamma}^{(m+1)}_P)^2 - 2\boldsymbol{\Gamma}^{(m+1)}_P\rho_0\boldsymbol{\Gamma}^{(m+1)}_P\Bigr) \\
&+ \mathcal{O}(\theta^3).
\end{aligned}
\]
The first term contains only \(\theta_j^2\) contributions by the induction hypothesis, while the second is proportional to \(\theta_{m+1}^2\). A cross term \(\theta_j\theta_{m+1}\) would require first-order contributions from both \(\rho_m\) and \(\mathcal{T}_{m+1}\), but both vanish. Therefore, \(\rho_{m+1}\) retains the same structure, completing the induction.

Therefore, by induction, the final state after all $n$ layers, $\rho_{\mathrm{RC}}=\mathcal{T}_n\circ\cdots\circ
\mathcal{T}_1(\rho_0)$, has the form
\begin{align}
\rho_{\mathrm{RC}}={}& \rho_0-\frac{1}{2}\sum_{j=1}^{n}\frac{\theta_j^2}{|\mathcal{P}|}\sum_{P\in\mathcal{P}}
\notag\\
&\Bigl((\boldsymbol{\Gamma}^{(j)}_P)^2\rho_0+\rho_0(\boldsymbol{\Gamma}^{(j)}_P)^2-2\boldsymbol{\Gamma}^{(j)}_P \rho_0\boldsymbol{\Gamma}^{(j)}_P \Bigr) +\mathcal{O}(\theta^3),
\label{RCdensitymatrix}
\end{align}
with no terms proportional to $\theta_j\theta_k$ for $j\neq k$.

\section{Details of Numerical Simulation in Sec.~\ref{section GBV representation}
\label{Appendix Simulation details}}

The density matrices $\rho_{\mathrm{bare}}$ and $\rho_{\mathrm{RC}}$ derived in Eqs.~(\ref{baredensitymatrix}) and~(\ref{RCdensitymatrix}) of Appendix~\ref{appendix RC-channel} are the starting points for the simulation reported in Sec.~\ref{section GBV representation}. Here we discuss further details of the simulation.

We perform numerical simulations using 1000 independent instances, each consisting of a random ideal final state $\rho_0$ and a random error configuration. $\rho_0$ is prepared as follows. A Haar-random unitary operator $U_{\mathrm{op}} = \prod_{j=n}^1 U_j^0$ in the full $2^N$-dimensional Hilbert space is sampled. The ideal output is then $\rho_0 = U_{\mathrm{op}} (\ket{0}\bra{0})^{\otimes N} U_{\mathrm{op}}^\dagger$, i.e., a Haar-random pure state in the Hilbert space. For each error configuration, the same $\rho_0$ is used for both the bare and RC evolutions, so that any difference in the final states can be attributed to the different error-accumulation mechanisms.

For the coherent error, as discussed in Appendix~\ref{appendix RC-channel}, we can directly generate the effective dressed error for each layer. A random $\boldsymbol{\Gamma}^{(j)}$ is constructed by conjugating a fixed diagonal matrix (normalized such that $\lambda_{\max}-\lambda_{\min}=2$, where $\lambda_{\max}$ and $\lambda_{\min}$ are its largest and smallest eigenvalues) with a Haar-random unitary. The error strength $\theta_j$ is drawn uniformly from $[-0.05,0.05]$. This construction ensures that the orientation of the error is uniformly random over the unitary group, while its strength is set solely by $\theta_j$. 

For the bare circuit, the evolution is determined by sequentially applying the erroneous layers; the final density matrix is uniquely determined once the error configuration is fixed. We apply RC to the same error configurations using two implementations: exhaustive and non‑exhaustive RC. Exhaustive RC was discussed in Appendix~\ref{appendix RC-channel}. For non‑exhaustive RC, we sample 1000 sequences uniformly from the \(4^{nN}\) possibilities, compute the output state for each, and average to obtain the final density matrix.

For each instance of a random $\rho_0$ and a random error configuration, we obtain three final density matrices for the bare, exhaustive RC, and non‑exhaustive RC circuits. For each type of circuit we have an ensemble of 1000 samples.

To understand why fidelity remains nearly unchanged while trace distance improves upon RC [Figs.~\ref{fig:GBVA}(e) and (f)], we examine the perturbative structure of $\rho_{\mathrm{bare}}$ and $\rho_{\mathrm{RC}}$ in Eqs.~(\ref{baredensitymatrix}) and~(\ref{RCdensitymatrix}).

For an ideal final state \(\rho_0 = |\psi\rangle\langle\psi|\), fidelity is \(F = \langle\psi|\rho_{\mathrm{out}}|\psi\rangle\). Substituting Eq.~(\ref{baredensitymatrix}) or Eq.~(\ref{RCdensitymatrix}) for $\rho_{\mathrm{out}}$ and using \(\langle\psi|[\boldsymbol{\Gamma}^{(j)},\rho_0]|\psi\rangle = 0\) gives \(F = 1 + \mathcal{O}(\theta^2)\) for both bare and exhaustive-RC circuits. Their layerwise second-order terms are related by Pauli averaging. Under the isotropic sampling used here, Pauli conjugation leaves the distribution of each $\boldsymbol{\Gamma}^{(j)}$ unchanged, while the inter-layer cross terms in the bare circuit have zero ensemble average because the traceless generators are independently oriented. The average fidelities of the bare and RC circuits are therefore nearly equal, as observed in Fig.~\ref{fig:GBVA}(e).

Trace distance is \(D(\rho_{\mathrm{out}},\rho_0) = \frac12 \|\rho_{\mathrm{out}} - \rho_0\|_1\). For the bare circuit, \(\|\rho_{\mathrm{bare}} - \rho_0\|_1 = \big\| i\sum_j \theta_j [\boldsymbol{\Gamma}^{(j)},\rho_0] \big\|_1 + \mathcal{O}(\theta^2)\), so the trace distance scales as \(\mathcal{O}(\theta)\). In the RC circuit the first‑order term is absent, giving \(\|\rho_{\mathrm{RC}} - \rho_0\|_1 = \mathcal{O}(\theta^2)\). Thus RC suppresses the trace distance from \(\mathcal{O}(\theta)\) to \(\mathcal{O}(\theta^2)\) while leaving fidelity essentially unchanged.
 
The simulation in Sec.~\ref{section GBV representation} shows a near-isotropic state-level contraction under coherent errors. This behavior does not follow from a single-layer Pauli twirl, which generally produces an anisotropic stochastic Pauli channel. Equation~(\ref{baredensitymatrix}) contains first-order coherent rotations and second-order cross terms between different layers. These cross terms are present whether or not the generators commute; their operator ordering records the noncommutative time ordering. By contrast, Eq.~(\ref{RCdensitymatrix}) shows that exhaustive RC removes the first-order terms and the inter-layer cross terms through second order, leaving a sum of layerwise second-order contributions.

This perturbative cancellation explains why RC suppresses coherent accumulation, but it does not by itself prove depolarizing behavior. Appendix~\ref{appendix convergence} derives the exact composition of the Pauli-transfer eigenvalues, and Appendix~\ref{appendix channel vector} gives a finite-depth, high-probability bound on their anisotropy under the independent, isotropically distributed weak coherent errors used in the simulation. The near-depolarizing statement therefore relies on this channel-space bound rather than on a generic random-walk or law-of-large-numbers argument.

\section{Composition of Multi-layer Stochastic Pauli Errors
\label{appendix convergence}}

In this appendix, we derive the exact structure of the exhaustive-RC
channel in the generalized Pauli basis. We first show explicitly how
the Pauli twirl converts a coherent error into a stochastic Pauli
channel, and then derive the action of the resulting multi-layer
channel on the GBV. The channel-space deviation from depolarization
and its high-probability bound are discussed in
Appendix~\ref{appendix channel vector}.

As in Appendix~\ref{appendix RC-channel}, we work in the effective
error frame in which the ideal evolution has already produced the
target state $\rho_0$. The coherent error in the $j$-th layer is
\begin{equation*}
U_j^e
=
\exp\left[-i\theta_j\boldsymbol{\Gamma}^{(j)}\right],
\end{equation*}
where $\boldsymbol{\Gamma}^{(j)}$ is a traceless Hermitian error
generator.

Because the $N$-qubit Pauli operators form a complete orthogonal
operator basis, $U_j^e$ can be expanded as
\begin{equation}
U_j^e
=
\sum_{Q\in\mathcal P}\alpha_j(Q)Q,
\qquad
\alpha_j(Q)
=
\frac{1}{d}
\operatorname{Tr}\left(Q^\dagger U_j^e\right),
\label{eq:appD-Pauli-expansion}
\end{equation}
where $\mathcal P$ is the $N$-qubit Pauli group with global phases
omitted. The orthogonality relation
\begin{equation*}
\operatorname{Tr}(Q^\dagger R)
=
d\,\delta_{Q,R},
\qquad Q,R\in\mathcal P,
\end{equation*}
together with the unitarity of $U_j^e$, gives
\begin{equation}
\sum_{Q\in\mathcal P}|\alpha_j(Q)|^2
=
\frac{1}{d}
\operatorname{Tr}\left[(U_j^e)^\dagger U_j^e\right]
=
1.
\label{eq:appD-alpha-normalization}
\end{equation}

For any two Pauli operators $P,Q\in\mathcal P$, conjugation by $P$
changes $Q$ by at most a sign:
\begin{equation*}
P^\dagger QP
=
\chi_Q(P)Q,
\qquad
\chi_Q(P)\in\{+1,-1\}.
\end{equation*}
Here $\chi_Q(P)=+1$ when $P$ and $Q$ commute and
$\chi_Q(P)=-1$ when they anticommute. It follows from
Eq.~(\ref{eq:appD-Pauli-expansion}) that
\begin{equation}
P^\dagger U_j^eP
=
\sum_{Q\in\mathcal P}
\alpha_j(Q)\chi_Q(P)Q.
\label{eq:appD-conjugated-error}
\end{equation}

The exhaustive Pauli twirl of the $j$-th layer is
\begin{equation}
\mathcal T_j(\rho)=\frac{1}{|\mathcal P|}\sum_{P\in\mathcal P}P^\dagger U_j^eP\,\rho\,P^\dagger(U_j^e)^\dagger P.
\label{eq:appD-twirl-definition}
\end{equation}
Substituting Eq.~(\ref{eq:appD-conjugated-error}) into Eq.~(\ref{eq:appD-twirl-definition}) gives
\begin{align}
\mathcal T_j(\rho)
&=
\frac{1}{|\mathcal P|}\sum_{P\in\mathcal P}\sum_{Q,R\in\mathcal P}\alpha_j(Q)\alpha_j(R)^*\chi_Q(P)\chi_R(P)
Q\rho R^\dagger
\notag\\
&=
\sum_{Q,R\in\mathcal P}\alpha_j(Q)\alpha_j(R)^*\left[\frac{1}{|\mathcal P|}\sum_{P\in\mathcal P}\chi_Q(P)\chi_R(P)
\right]Q\rho R^\dagger.
\label{eq:appD-twirl-expanded}
\end{align}
Notice that $\chi_Q(P)\chi_R(P)$ is precisely the sign acquired by $QR$ under conjugation by $P$, since
\begin{equation*}
P^\dagger QRP = P^\dagger QPP^\dagger RP = \chi_Q(P)\chi_R(P)QR.
\end{equation*}
If $Q\neq R$ (with global phases omitted), then $QR$ is a nonidentity Pauli operator. Exactly half of the Pauli operators $P$ commute with $QR$ and half anticommute with it. Hence $\chi_Q(P)\chi_R(P)$ equals $+1$ for half of the terms and $-1$ for the other half, so their sum vanishes. If $Q=R$, then $\chi_Q(P)\chi_R(P)=\chi_Q(P)^2=1$ for every $P$. Therefore, the Pauli characters satisfy
\begin{equation}
\frac{1}{|\mathcal P|}\sum_{P\in\mathcal P}\chi_Q(P)\chi_R(P)=\delta_{Q,R}.
\label{eq:appD-character-orthogonality}
\end{equation}

Equation~(\ref{eq:appD-character-orthogonality}) eliminates all off-diagonal terms with $Q\neq R$ in
Eq.~(\ref{eq:appD-twirl-expanded}). Therefore, 
\begin{equation}
\mathcal T_j(\rho)=\sum_{Q\in\mathcal P}|\alpha_j(Q)|^2Q\rho Q^\dagger=\sum_{Q\in\mathcal P}p_j(Q)Q\rho Q^\dagger,
\label{eq:appD-layer-channel}
\end{equation}
where
\begin{equation*}
p_j(Q)=|\alpha_j(Q)|^2.
\end{equation*}
Equation~(\ref{eq:appD-alpha-normalization}) ensures that $\sum_{Q\in\mathcal P}p_j(Q)=1.$ Thus the exhaustive Pauli twirl converts the coherent error in each layer exactly into a stochastic Pauli channel. In general, however,
the probabilities $p_j(Q)$ are not equal, so a single twirled layer need not be depolarizing.

The exhaustive-RC channel for the complete $n$-layer circuit is the composition
\begin{equation*}
\mathcal T_{\mathrm{RC}}=\mathcal T_n\circ\mathcal T_{n-1}\circ\cdots\circ\mathcal T_1.
\end{equation*}
Expanding this composition gives
\begin{align*}
\mathcal T_{\mathrm{RC}}(\rho)
&=
\sum_{Q_1,\ldots,Q_n\in\mathcal P}\left[\prod_{j=1}^{n}p_j(Q_j)\right](Q_n\cdots Q_1)\rho(Q_n\cdots Q_1)^\dagger
\notag\\
&=
\sum_{S\in\mathcal P}p_{\mathrm{RC}}(S)S\rho S^\dagger,
\end{align*}
where
\begin{equation*}
p_{\mathrm{RC}}(S)=\sum_{\substack{Q_1,\ldots,Q_n\in\mathcal P\\Q_n\cdots Q_1=S}}\prod_{j=1}^{n}p_j(Q_j).
\end{equation*}
This is the exact convolution of the layer-wise Pauli distributions and does not require the different layers to be identically distributed.

To connect this result with the GBV representation, let $\Gamma_a$, $a=1,\ldots,d^2-1$, denote the generalized Pauli matrices introduced in Eq.~(\ref{GBVdefinition}). For any
$Q\in\mathcal P$,
\begin{equation*}
Q\Gamma_aQ^\dagger = \chi_a(Q)\Gamma_a,\qquad \chi_a(Q)\in\{+1,-1\}.
\end{equation*}
Combining Eq.~(\ref{eq:appD-layer-channel}) with the generalized Pauli expansion of a density matrix in Eq.~(\ref{GBVdefinition}), we can evaluate the action of $\mathcal T_j$ separately on the identity and the traceless generalized Pauli components. By linearity, $\mathcal T_j(I)=I$, while for each nonidentity generalized Pauli
matrix $\Gamma_a$,
\begin{equation*}
\mathcal T_j(\Gamma_a)=\sum_{Q\in\mathcal P}p_j(Q)Q\Gamma_aQ^\dagger = \left[\sum_{Q\in\mathcal P}p_j(Q)\chi_a(Q)
\right]\Gamma_a.
\end{equation*}
Defining $\lambda_a^{(j)}=\sum_{Q\in\mathcal P}p_j(Q)\chi_a(Q)$, we obtain
\begin{equation*}
\mathcal T_j(\Gamma_a) = \lambda_a^{(j)}\Gamma_a.
\end{equation*}
Thus each $\Gamma_a$ is an eigenoperator of $\mathcal T_j$, and the action of every layer channel on the traceless operator subspace is diagonal in the generalized Pauli basis.

Because the layer channels are diagonal in the same basis on this subspace, their composition satisfies
\begin{equation}
\mathcal T_{\mathrm{RC}}(\Gamma_a)=\lambda_a^{\mathrm{RC}}\Gamma_a,\qquad\lambda_a^{\mathrm{RC}}
=\prod_{j=1}^{n}\lambda_a^{(j)}.
\label{eq:appD-total-eigenvalue}
\end{equation}
Accordingly, the action of the complete exhaustive-RC channel on the GBV components is represented by the $(d^2-1)\times(d^2-1)$ diagonal matrix
\begin{equation}
R_{\mathrm{RC}}=\operatorname{diag}\left(\lambda_1^{\mathrm{RC}},\ldots,\lambda_{d^2-1}^{\mathrm{RC}}\right).
\label{eq:appD-total-PTM}
\end{equation}

Writing the ideal output state as 
\begin{equation*}
\rho_0=\frac{I}{d}+\frac{1}{d}\sum_{a=1}^{d^2-1}r_a^0\Gamma_a,
\end{equation*}
the exhaustive-RC output state is
\begin{equation*}
\rho_{\mathrm{RC}} = \mathcal T_{\mathrm{RC}}(\rho_0) = \frac{I}{d}+\frac{1}{d}\sum_{a=1}^{d^2-1}
\lambda_a^{\mathrm{RC}}r_a^0\Gamma_a.
\end{equation*}
Equivalently, the GBV transforms as
\begin{equation*}
\vec r_{\mathrm{RC}}
=
R_{\mathrm{RC}}\vec r_0.
\end{equation*}
Therefore, each GBV component is contracted by the corresponding Pauli-transfer eigenvalue $\lambda_a^{\mathrm{RC}}$.

For comparison, define the depolarizing channel
\begin{equation}
\mathcal D_p^{\mathrm{dep}}(X)=pX+(1-p)\operatorname{Tr}(X)\frac{I}{d}.
\label{eq:appD-depolarizing-channel}
\end{equation}
Its action on every traceless generalized Pauli matrix is
\begin{equation*}
\mathcal D_p^{\mathrm{dep}}(\Gamma_a)=p\Gamma_a.
\end{equation*}
Consequently,
\begin{equation*}
\mathcal T_{\mathrm{RC}}=\mathcal D_p^{\mathrm{dep}}~\Longleftrightarrow~\lambda_1^{\mathrm{RC}}
=\lambda_2^{\mathrm{RC}}=\cdots=\lambda_{d^2-1}^{\mathrm{RC}}=p.
\end{equation*}
When this condition holds,
\begin{equation*}
\vec r_{\mathrm{RC}}=p\vec r_0,
\end{equation*}
so the channel contracts the GBV isotropically without changing its orientation.

Therefore, the deviation of $\mathcal T_{\mathrm{RC}}$ from a depolarizing channel is determined by the dispersion among the Pauli-transfer eigenvalues $\lambda_a^{\mathrm{RC}}$. Appendix~\ref{appendix channel vector} introduces a channel-space angle that quantifies this dispersion and gives a high-probability bound under the independent, isotropically distributed weak coherent errors considered in this work.

\begin{figure}[htbp]
    \centering
    \includegraphics[scale=0.34]{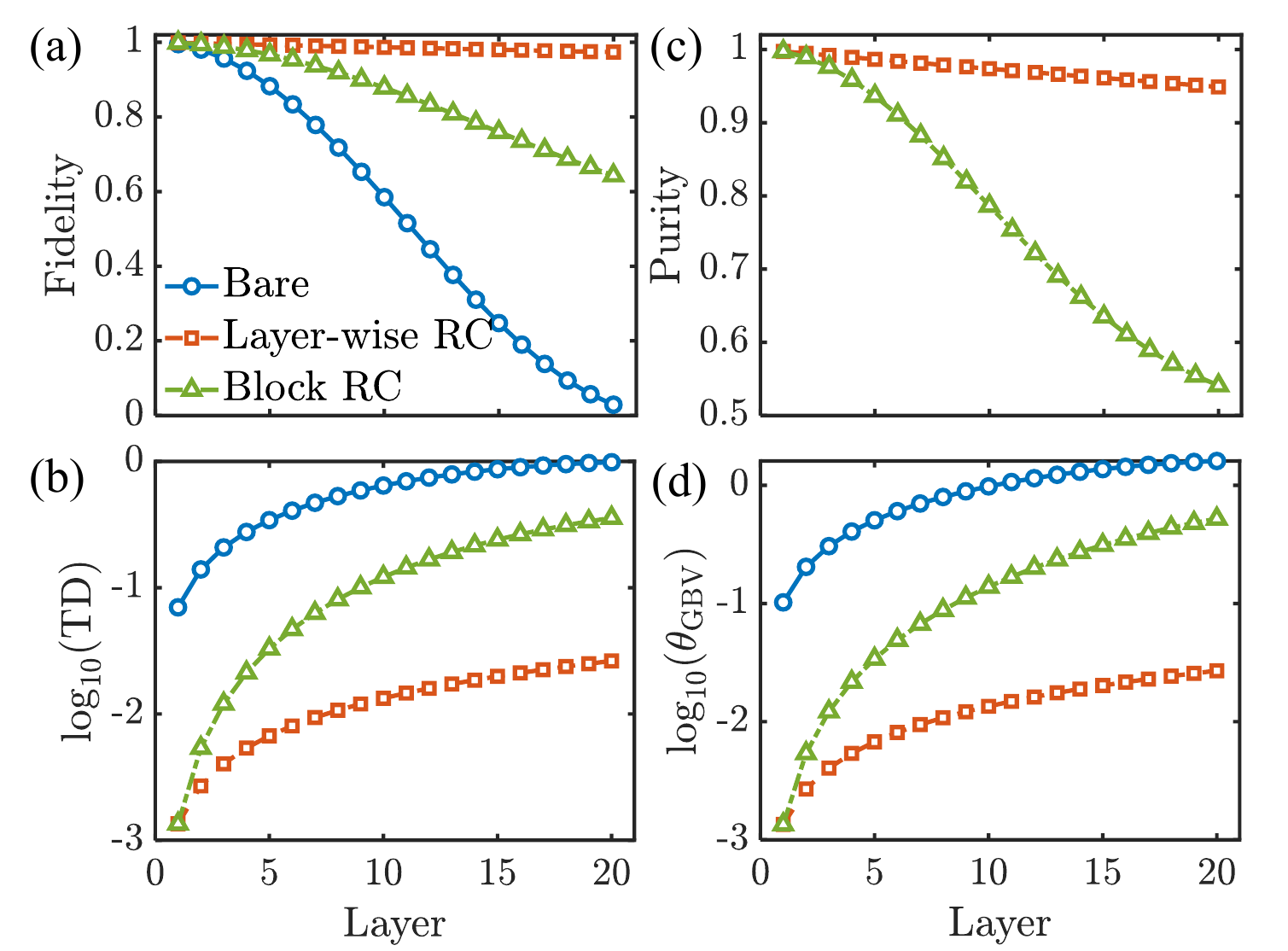}
    \caption{Block-RC under perfectly correlated coherent errors. Fidelity, trace distance, purity, and GBV angle are compared for bare evolution, layer-wise RC, and single-block RC as functions of circuit depth $n$. In single-block RC, all $n$ layers form one block, with randomization only at its boundaries. Both RC protocols use exhaustive Pauli averaging.}
    \label{fig:block_rc_correlated_errors}
\end{figure}

\section{Block-RC Under Perfectly Correlated Coherent Errors}
\label{appendix Block-RC}

To illustrate the limitation discussed in Sec.~\ref{Block-RC}, we consider the following example. The ideal output is a four-qubit GHZ state, $\ket{\mathrm{GHZ}}=(\ket{0000}+\ket{1111})/\sqrt{2}$. The simulation is performed in the toggling frame, starting directly from this ideal output state. Each layer applies the same effective coherent error,
\[
\begin{gathered}
U_j^e=U_e=\bigotimes_{q=1}^{4}\operatorname{diag}(1,e^{i\phi_q}),\\
(\phi_1,\phi_2,\phi_3,\phi_4)=(0.02,0.03,0.04,0.05)~\mathrm{rad}.
\end{gathered}
\]
Thus the errors are perfectly correlated across layers. We compare independent layer-wise Pauli twirling with a single Pauli twirl of the entire $n$-layer block. Both averages are exhaustive, and the inserted Pauli operations are ideal.

Figure~\ref{fig:block_rc_correlated_errors} shows that single-block RC produces a larger GBV angle and trace distance, together with lower fidelity and purity, than layer-wise RC. The differences become more pronounced as the depth increases. At $n=20$, the fidelity and purity are $0.974$ and $0.949$ for layer-wise RC, compared with $0.643$ and $0.541$ for single-block RC.

This difference arises from coherent accumulation within the block. The identical phase errors add in every layer, giving a total phase $n\phi_q$ on each qubit. Randomization at the block boundaries does not interrupt this accumulation, whereas layer-wise RC independently randomizes the phase signs at each layer. Here, increasing the circuit depth also increases the block size, since all $n$ layers form one block. The comparable performance of block-RC and layer-wise RC under independent random errors in Fig.~\ref{fig:GBVA} therefore need not hold under strong inter-layer correlations.

\section{Channel-Space Quantification of Near-Depolarizing Behavior
\label{appendix channel vector}}

The Pauli-basis representation derived in Appendix~\ref{appendix convergence} provides a direct way to quantify the term ``near-depolarizing.'' We define a channel-space angle relative to the depolarizing direction and derive a finite-depth, high-probability bound on this angle for independent and isotropically distributed weak coherent errors.

Define the normalized operator basis
\begin{equation*}
\widetilde{\Gamma}_0=\frac{I}{\sqrt d},
\qquad
\widetilde{\Gamma}_a=\frac{\Gamma_a}{\sqrt d},
\quad
a=1,\ldots,d^2-1,
\end{equation*}
which satisfies
\begin{equation*}
\operatorname{Tr}
\left(
\widetilde{\Gamma}_\mu^\dagger
\widetilde{\Gamma}_\nu
\right)
=
\delta_{\mu\nu},
\qquad
\mu,\nu=0,\ldots,d^2-1.
\end{equation*}
All channels considered here are trace-preserving and unital. Their
action on the traceless operator subspace is described by the reduced
Pauli-transfer matrix
\begin{equation*}
R_{ab}(\mathcal E)
=
\operatorname{Tr}
\left[
\widetilde{\Gamma}_a
\mathcal E(\widetilde{\Gamma}_b)
\right]
=
\frac{1}{d}
\operatorname{Tr}
\left[
\Gamma_a\mathcal E(\Gamma_b)
\right],
\end{equation*}
where $a,b=1,\ldots,d^2-1$.

The completely depolarizing channel
\begin{equation*}
\mathcal D_0^{\mathrm{dep}}(X)
=
\operatorname{Tr}(X)\frac{I}{d}
\end{equation*}
has the same action as $\mathcal E$ on the identity component. Therefore, subtracting $\mathcal D_0^{\mathrm{dep}}$ removes the fixed identity component, leaving the coefficients $R_{ab}(\mathcal E)$. We define the corresponding channel vector as
\begin{equation*}
\vec c_{\mathrm{ch}}(\mathcal E)
\coloneqq
\operatorname{vec}[R(\mathcal E)]
\in
\mathbb R^{(d^2-1)^2}.
\end{equation*}
The stacking convention in $\operatorname{vec}[R]$ only permutes the vector components and therefore does not affect inner products or angles. Likewise, an orthonormal change of operator basis transforms
$R\mapsto ORO^T$, leaving both $\operatorname{Tr}(R)$ and the Frobenius norm $\|R\|_{\mathrm F}\coloneqq
\sqrt{\operatorname{Tr}(R^\dagger R)}$ invariant.

For the depolarizing channel defined in
Eq.~(\ref{eq:appD-depolarizing-channel}),
\begin{equation*}
R(\mathcal D_p^{\mathrm{dep}})
=
pI_{d^2-1}.
\end{equation*}
Thus all depolarizing channels with $p>0$ lie along the direction
$\operatorname{vec}(I_{d^2-1})$. The projection of a general nonzero $R$ onto
this direction is $p_*I_{d^2-1}$, where
\begin{equation*}
p_*
=
\frac{\operatorname{Tr}(R)}{d^2-1}.
\end{equation*}
For $R\neq0$, we define the channel-space angle by
\begin{equation*}
\theta_{\mathrm{ch}}(\mathcal E)
=
\arccos
\left[
\frac{\operatorname{Tr}(R)}
{\sqrt{d^2-1}\,\|R\|_{\mathrm F}}
\right].
\end{equation*}
For $p_*>0$, the channel is depolarizing if and only if
$\theta_{\mathrm{ch}}=0$. The completely depolarizing channel is identified separately by $R=0$.

Writing
\begin{equation*}
R=p_*I_{d^2-1}+R_\perp,
\qquad
\operatorname{Tr}(R_\perp)=0,
\end{equation*}
gives the equivalent expression
\begin{equation*}
\tan\theta_{\mathrm{ch}}(\mathcal E)
=
\frac{\|R_\perp\|_{\mathrm F}}
{\sqrt{d^2-1}\,p_*}.
\end{equation*}
Thus $\tan\theta_{\mathrm{ch}}$ is the relative magnitude of the
anisotropic part of the channel.

For exhaustive RC in the weak-error regime considered here, using Eq.~(\ref{eq:appD-total-PTM}), we obtain
\begin{equation}
\tan\theta_{\mathrm{ch}}(\mathcal T_{\mathrm{RC}}) = \frac{\left[\frac{1}{d^2-1}\sum_{a=1}^{d^2-1}\left(
\lambda_a^{\mathrm{RC}}-\overline{\lambda}\right)^2\right]^{1/2}}{\overline{\lambda}},
\label{eq:appE-Pauli-angle}
\end{equation}
where $\overline{\lambda}=\frac{1}{d^2-1}\sum_{a=1}^{d^2-1}\lambda_a^{\mathrm{RC}}.$
Equation~(\ref{eq:appE-Pauli-angle}) is the quantitative criterion used here for near-depolarizing behavior.

We now derive a high-probability bound for
$\theta_{\mathrm{ch}}$. As in Appendix~
\ref{Appendix Simulation details}, we condition on the error
strengths $\{\theta_j\}$ and write
\begin{equation*}
\boldsymbol{\Gamma}^{(j)}
=
V_j\boldsymbol{\Gamma}_{\mathrm{ref}}V_j^\dagger,
\end{equation*}
where the $V_j$ are independent Haar-random unitaries and
$\boldsymbol{\Gamma}_{\mathrm{ref}}$ is a fixed traceless Hermitian
matrix with spectral range two. Hereafter, $\mathbb E$ and
$\mathbb P$ denote expectation and probability over the random
orientations $\{V_j\}$ at fixed $\{\theta_j\}$.

The Pauli-transfer eigenvalue of the $j$-th twirled layer is
\begin{align*}
\lambda_a^{(j)}
&=
\frac{1}{d}
\operatorname{Tr}
\left[
\Gamma_aU_j^e\Gamma_a(U_j^e)^\dagger
\right]
\\
&=
\frac{1}{d}
\sum_{k,l}
|(\Gamma_a)_{kl}|^2
\cos
\left[
\theta_j
\left(
g_k^{(j)}-g_l^{(j)}
\right)
\right],
\end{align*}
where $g_k^{(j)}$ are the eigenvalues of
$\boldsymbol{\Gamma}^{(j)}$, and $(\Gamma_a)_{kl}$ are the matrix elements of $\Gamma_a$ in the corresponding eigenbasis. Because
$\sum_{k,l}|(\Gamma_a)_{kl}|^2=d$ and
$|g_k^{(j)}-g_l^{(j)}|\le2$, we have
\begin{equation*}
\cos(2|\theta_j|)
\le
\lambda_a^{(j)}
\le
1.
\end{equation*}

According to Eq.~(\ref{eq:appD-total-eigenvalue}), the accumulated eigenvalue $\lambda_a^{\mathrm{RC}}$ is a product of the layer-wise eigenvalues. To convert this product into a sum, for $|\theta_j|<\pi/4$ we define
\begin{equation*}
X_{j,a}=-\log\lambda_a^{(j)},~~0\le X_{j,a}\le b_j,~~b_j=-\log[\cos(2|\theta_j|)].
\end{equation*}
Haar invariance implies that the expectation $m_j=\mathbb E[X_{j,a}]$ is independent of the Pauli direction $a$. We separate the common mean attenuation from the direction-dependent fluctuation by defining
\begin{equation*}
\mu=\sum_{j=1}^{n}m_j,~~Z_a=\sum_{j=1}^{n}(X_{j,a}-m_j).
\end{equation*}
The accumulated eigenvalues can then be written as
\begin{equation}
\lambda_a^{\mathrm{RC}}=e^{-\mu-Z_a}.
\label{eq:appE-total-eigenvalue}
\end{equation}
Here $\mu$ is common to all Pauli directions, whereas $Z_a$ describes the deviation of direction $a$ from this common attenuation. Therefore, showing that all $Z_a$ are small is sufficient to show that the accumulated eigenvalues are nearly equal.

To bound these fluctuations, we use Hoeffding's inequality, which controls the deviation of a sum of independent bounded random variables from its expectation. For each fixed $a$, the variables $X_{j,a}$ are independent across
layers and bounded by $0\le X_{j,a}\le b_j$. Hoeffding's inequality
therefore gives
\begin{equation*}
\mathbb P(|Z_a|\ge t)\le2\exp\left(-\frac{2t^2}{\sum_{j=1}^{n}b_j^2}\right).
\end{equation*}

Let $0<\delta<1$ be a chosen failure probability, so that $1-\delta$ is the corresponding confidence level. Define
\begin{equation}
t_\delta = \left(\sum_{j=1}^{n}b_j^2\right)^{1/2}\sqrt{\frac{1}{2}\log\left[\frac{2(d^2-1)}{\delta}\right]}.
\label{eq:appE-t-delta}
\end{equation}
Applying the union bound to all $d^2-1$ Pauli directions then gives
\begin{equation*}
\mathbb P\left(\max_a|Z_a|\le t_\delta\right)\ge1-\delta.
\end{equation*}

On this event, Eq.~(\ref{eq:appE-total-eigenvalue}) gives
\begin{equation*}
e^{-\mu-t_\delta} \le \lambda_a^{\mathrm{RC}} \le e^{-\mu+t_\delta}
\end{equation*}
for every $a$. It follows that
\begin{equation*}
\overline{\lambda} \ge e^{-\mu-t_\delta},
\end{equation*}
and
\begin{equation*}
\left[
\frac{1}{d^2-1}
\sum_{a=1}^{d^2-1}
\left(
\lambda_a^{\mathrm{RC}}-\overline{\lambda}
\right)^2
\right]^{1/2}
\le
e^{-\mu}\left(e^{t_\delta}-1\right).
\end{equation*}
Substituting these two bounds into
Eq.~(\ref{eq:appE-Pauli-angle}) yields
\begin{equation}
\tan\theta_{\mathrm{ch}}(\mathcal T_{\mathrm{RC}})
\le
e^{t_\delta}\left(e^{t_\delta}-1\right)
\label{eq:appE-channel-angle-bound}
\end{equation}
with probability at least $1-\delta$. This is an input-independent,
finite-depth bound on the deviation from the depolarizing direction.

For the parameters used in Fig.~\ref{fig:GBVA}, $d=16$, $n=20$, and $|\theta_j|\le0.05$. Hence, $b_j \le -\log[\cos(0.1)] \approx 0.00501$ and $\left(\sum_{j=1}^{20}b_j^2\right)^{1/2} \le 0.02240.$ Taking $\delta=0.05$ in
Eq.~(\ref{eq:appE-t-delta}) gives $t_\delta\le0.04812$. Equation~(\ref{eq:appE-channel-angle-bound}) then yields $\tan\theta_{\mathrm{ch}}(\mathcal T_{\mathrm{RC}}) \le 0.05173$, or equivalently, $\theta_{\mathrm{ch}}(\mathcal T_{\mathrm{RC}}) \le 0.05168~\mathrm{rad}$, with probability at least $95\%$. This provides a quantitative
channel-level meaning of ``near-depolarizing'' for the parameter range studied in Fig.~\ref{fig:GBVA}.

\section{Finite-Sampling Error of Non-exhaustive RC \label{appendix non-EX RC}}

In this Appendix, we quantify the error incurred when the exhaustive-RC average over all $4^{nN}$ sequences is approximated using $M$ independently sampled sequences. The result follows from elementary probability theory and the boundedness of the quantum state space.

Consider a fixed coherent-error configuration in the coherent-error-only setting of Sec.~\ref{Simulation}. For $M$ independent random RC sequences, define the error vector of the $k$-th sequence as
\[
\mathbf{e}_{k} = \mathbf{r}_{\text{RC}}^{(k)} - \mathbf{r}_{\text{ideal}},
\]
where $\mathbf{r}_{\text{RC}}^{(k)}$ and $\mathbf{r}_{\text{ideal}}$ are the GBVs of the corresponding RC output and the ideal output, respectively. Because the sequences are generated independently using the same randomization rule, the vectors $\mathbf{e}_{1},\ldots,\mathbf{e}_{M}$ are independent and identically distributed.

Define the exhaustive-RC average and the sample average over $M$ sequences as
\[
\bar{\mathbf{e}} = \mathbb{E}[\mathbf{e}_{k}],
\qquad
\hat{\mathbf{e}} = \frac{1}{M}\sum_{k=1}^{M}\mathbf{e}_{k}.
\]
Independence eliminates the cross terms between different centered error vectors, giving
\[
\begin{aligned}
\mathbb{E}[|\hat{\mathbf{e}}-\bar{\mathbf{e}}|^{2}]
&= \frac{1}{M}\mathbb{E}[|\mathbf{e}_{1}-\bar{\mathbf{e}}|^{2}]\\
&= \frac{1}{M}\left(\mathbb{E}[|\mathbf{e}_{1}|^{2}]-|\bar{\mathbf{e}}|^{2}\right)
\le \frac{\mathbb{E}[|\mathbf{e}_{1}|^{2}]}{M}.
\end{aligned}
\]

The bounded GBV length also gives an absolute estimate. For any physical $N$-qubit state, $|\mathbf{r}|\le\sqrt{2^N-1}$. The triangle inequality therefore gives
\[
|\mathbf{e}_{1}|
\le |\mathbf{r}_{\mathrm{RC}}^{(1)}|+|\mathbf{r}_{\mathrm{ideal}}|
\le 2\sqrt{2^N-1},
\]
and hence
\[
\mathbb{E}[|\hat{\mathbf{e}}-\bar{\mathbf{e}}|^{2}]
\le \frac{4(2^N-1)}{M}.
\]

More importantly, the relative root-mean-square sampling error satisfies
\[
\frac{\sqrt{\mathbb{E}[|\hat{\mathbf{e}}-\bar{\mathbf{e}}|^{2}]}}
{\sqrt{\mathbb{E}[|\mathbf{e}_k|^{2}]}}
=
\frac{1}{\sqrt{M}}
\left(
1-\frac{|\bar{\mathbf{e}}|^{2}}
{\mathbb{E}[|\mathbf{e}_1|^{2}]}
\right)^{1/2}
\le
\frac{1}{\sqrt{M}}.
\]
This normalized bound is independent of the system size and scales as $M^{-1/2}$. For the $M=1000$ sequences used in our simulations, the upper bound is approximately $0.0316$. Consistently, Fig.~\ref{fig:GBVA} shows that the purity and $\theta_{\mathrm{GBV}}$ obtained with non-exhaustive RC are visually indistinguishable from the exhaustive-RC results.

The argument above controls only the sampling error $\hat{\mathbf{e}}-\bar{\mathbf{e}}$; it does not determine the magnitude or orientation of the exhaustive-RC average $\bar{\mathbf{e}}$. Those properties are analyzed in Appendices~\ref{appendix RC-channel}--\ref{appendix channel vector}. The $M^{-1/2}$ bound should therefore be interpreted as a convergence guarantee for finite sampling, not as evidence of near-depolarizing behavior.

\section{Experimental Setup and Device Parameters \label{appendix Exp-setup}}

\begin{figure}[H]
    \centering
    \includegraphics[scale=0.22]{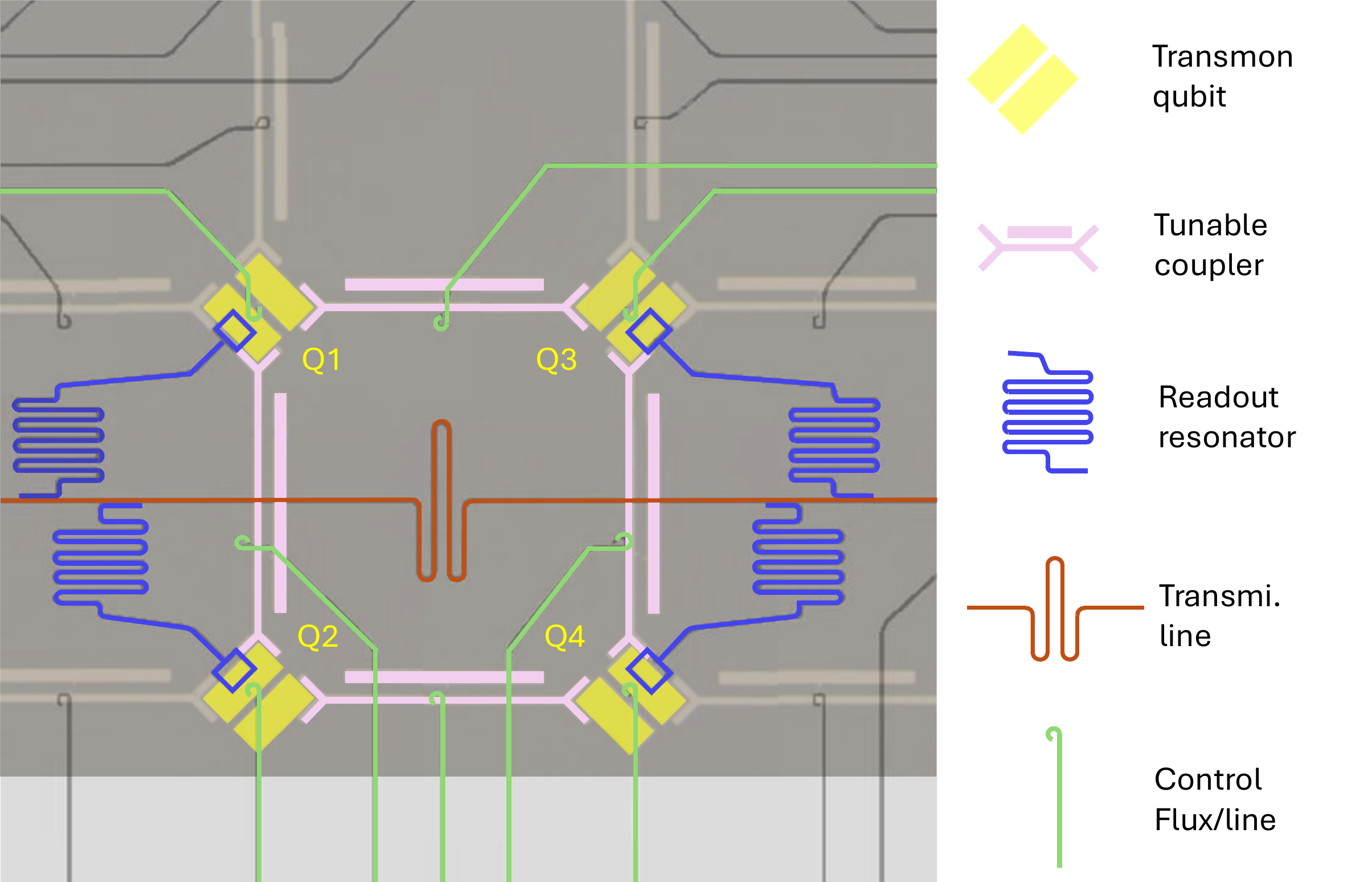}
    \caption{The superconducting quantum chip used in the experiment. The four qubits used in this work are highlighted in yellow.}
    \label{fig:chip}
\end{figure}

The superconducting quantum processor used in our experiment adopts a flip-chip packaging architecture. The top chip contains only the qubits and couplers, while the bottom chip hosts all the control lines, transmission lines, and readout resonators. The entire processor comprises a $4 \times 4$ array of frequency-tunable transmon qubits. In our experiment, we selected four qubits arranged in a closed-loop configuration, as shown in Fig.~\ref{fig:chip}. The coupling between qubits is mediated by tunable couplers~\cite{Tunablb2018Yan}. High-frequency microwave signals are used to drive the qubits and implement high-fidelity single-qubit gates, whereas low-frequency signals are employed to tune the frequencies of qubits and couplers, effectively controlling the inter-qubit coupling strength and enabling the realization of two-qubit gates such as the iSWAP and CZ gates. Each qubit is capacitively coupled to a dedicated readout resonator. A complete list of the key system parameters—including qubit frequencies, $T_1$, $T_2$, and readout fidelities—is provided in Table~\ref{table1}.

\begin{table}[H]
\centering
\caption{Device parameters of the four qubits used in the experiment.}
\label{table1}
\begin{tabular}{ c | c | c | c | c }
  \hline
  Qubits & Q1 & Q2 & Q3 & Q4 \\ \hline
  Frequency (GHz) & 4.385 & 4.061 & 4.102 & 4.350 \\ 
  Anharmonicity (MHz) & 175 & 182 & 190 & 182 \\
  Resonator frequency (GHz) & 6.489 & 6.414 & 6.370 & 6.617 \\
  Resonator linewidth (MHz) & 1.7 & 1.1 & 1.0 & 1.1 \\
  Dispersive shift (MHz)  & 1.1 & 0.7 & 1.0 & 1.0 \\
  Readout fidelity of $\ket{0}$ (\%) & 89.8 & 95.9 & 90.7 & 94.6 \\
  Readout fidelity of $\ket{1}$ (\%) & 85.6 & 92.6 & 88.8 & 89.7 \\
  Relaxation time of $|1\rangle$ (µs) & 36.87 & 22.91 & 27.53 & 31.61\\
  Ramsey decay time (µs) & 5.45 & 3.00 & 3.88 & 8.35 \\
  Spin echo decay time (µs) & 10.47 & 14.38 & 10.77 & 21.51 \\
  \hline
\end{tabular}
\end{table}

\begin{figure}[htbp]
    \centering
    \includegraphics[scale=0.62]{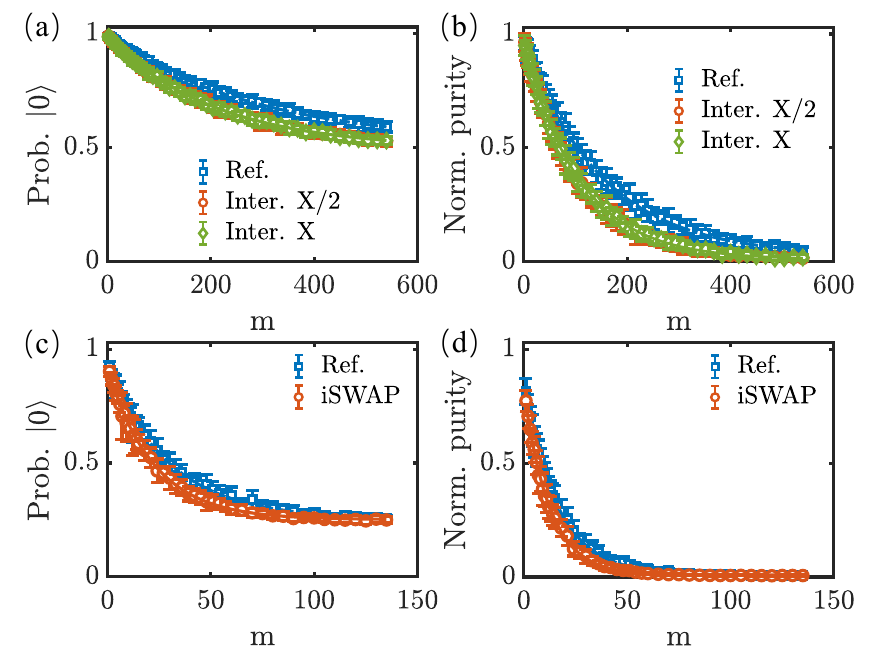}
    \caption{RB and purity-RB results for single‑qubit and iSWAP gates. Fidelities from RB are $0.99930$ ($X/2$), $0.99927$ ($X$), and $0.9917$ (iSWAP). Incoherent fractions from purity RB are $98.2\%$ ($X/2$), $94.6\%$ ($X$), and $84.0\%$ (iSWAP).}
    \label{fig:rb}
\end{figure}

\section{Further Data Analysis for GHZ-state Preparation Experiment \label{appendix GHZ analysis}}

Before applying any mitigation, we first establish an initial error budget using gate‑level benchmarking. Standard randomized benchmarking (RB) quantifies total gate infidelity; purity-RB measures the normalized purity $P_{\mathrm{normalized}} = \frac{d}{d-1}\left[\operatorname{Tr}(\rho^2) - \frac{1}{d}\right]$ ($d$ is the Hilbert space dimension) to extract the incoherent error fraction.

Interleaved RB yields the gate fidelity, while purity-RB yields the incoherent fraction. As shown in Figs.~\ref{fig:rb}(a) and (b), single‑qubit gates have fidelities $>0.999$ with incoherent errors accounting for more than $94\%$ of the gate infidelity. The two‑qubit iSWAP gate (Figs.~\ref{fig:rb}(c) and (d)) has fidelity $0.9917$, with $84\%$ of the gate infidelity attributed to incoherent errors. Although the coherent fraction is small, the analysis in Sec.~\ref{section GBV representation} suggests that such errors can induce orientational deviations that grow with circuit depth—an effect not captured by fidelity metrics alone.

\begin{table}[H]
\centering
\caption{Angles between GBVs.}
\label{tab:GBV-angle}
\begin{tabular}{lccccc}
\hline
 & $\rho_{\mathrm{ideal}}$ & $\rho_{\mathrm{bare}}$ & $\rho_{\mathrm{RC}}$ & $\rho_{\mathrm{bare}}^i$ & $\rho_{\mathrm{RC}}^i$\\
\hline
$\rho_{\mathrm{ideal}}$   & 0 & 0.261 & 0.138 & 0.281$\pm$0.013 & 0.265$\pm$0.036 \\
$\rho_{\mathrm{bare}}$    & 0.261 & 0 & 0.222 & / & / \\
$\rho_{\mathrm{RC}}$      & 0.138 & 0.222 & 0 & / & / \\
$\rho_{\mathrm{bare}}^j$  & 0.281$\pm$0.013 & / & / & 0.107$\pm$0.059 & / \\
$\rho_{\mathrm{RC}}^j$    & 0.265$\pm$0.036 & / & / & / & 0.337$\pm$0.106 \\
\hline
\end{tabular}
\end{table}

Table~\ref{tab:GBV-angle} summarizes the GBV angles for both individual realizations and their ensemble averages in the GHZ-state preparation experiment. The mean angle between individual RC sequences and the ideal state, \(\theta_{\mathrm{GBV}}(\rho_{\mathrm{ideal}},\rho_{\mathrm{RC}}^i)=0.265\), is only slightly smaller than that of bare individual sequences (0.281). Yet the ensemble‑averaged RC state lies at 0.138 from the ideal—a marked improvement. 

This contrast illustrates a key feature of randomization: individual RC sequences deviate in different orientations, so their average points much closer to the ideal. The increased spread under RC also appears in the angles between individual sequences: \(\theta_{\mathrm{GBV}}(\rho_{\mathrm{RC}}^i,\rho_{\mathrm{RC}}^j)=0.337\pm0.106\), compared to \(0.107\pm0.059\) for the bare circuit. The angle between the bare and RC averaged states, \(\theta_{\mathrm{GBV}}(\rho_{\mathrm{bare}},\rho_{\mathrm{RC}})=0.222\), shows that RC does not simply pull the state toward the ideal in a coplanar way; instead, it reorients the state in the high‑dimensional GBV space.

\section{The $D(\mathbb{Z}_3)$ Quantum Double Model and Anyon Braiding \label{appendix braiding theory}}

The quantum phase estimation experiment discussed in Sec.~\ref{section braiding experiment} simulates the braiding of anyons in the $D(\mathbb{Z}_3)$ quantum double model (also known as the $\mathbb{Z}_3$ toric code). This Appendix briefly outlines the theoretical background of this model and explains how it is mapped onto the quantum circuit implemented in our experiment.

In topological quantum phases, quasiparticles called anyons obey exotic statistics: exchanging two anyons can transform the system's state in more complex ways than simply acquiring a sign. This leads to two classes: Abelian anyons, where exchange yields a global phase factor, and non-Abelian anyons, where it performs a rotation within a degenerate state space. These exchange operations are physically realized by braiding—adiabatically moving anyons around one another so that their worldlines intertwine in spacetime. 

A canonical family of models describing such topological phases is Kitaev's quantum double model. Within this family, the $D(\mathbb{Z}_3)$ model serves as a prototypical example of a two-dimensional Abelian topological order. Its anyons are labeled by a pair of indices $(e^g, m^h)$. Here, $g$ and $h$ are elements of the cyclic group $\mathbb{Z}_3 = \{0, 1, 2\}$, denoting the electric charge and magnetic flux carried by the anyon, respectively. Consequently, the model hosts 9 distinct anyon types. The trivial vacuum is denoted as $(e^0, m^0)$. Nontrivial excitations include pure charges such as $(e^1, m^0)$ and pure fluxes like $(e^0, m^2)$, and composite particles known as dyons where both indices are non-zero (e.g., $(e^1, m^2)$). 

\begin{table*}[ht]
\centering
\caption{Nonzero coefficients in the Pauli expansion of 
$\ket{\psi_{e^g}}\bra{1111}=\sum_j p_j \Gamma_j$.}
\begin{tabular}{cc|cc|cc|cc|cc|cc}
\hline
$\Gamma_j$ & $p_j$ & $\Gamma_j$ & $p_j$ & $\Gamma_j$ & $p_j$ 
& $\Gamma_j$ & $p_j$ & $\Gamma_j$ & $p_j$ & $\Gamma_j$ & $p_j$ \\
\hline

IXIX & $-\frac{\sqrt3}{96}-\frac{i}{32}$ &
IXIY & $-\frac{1}{32}+\frac{i\sqrt3}{96}$ &
IXZX & $\frac{\sqrt3}{96}+\frac{i}{32}$ &
IXZY & $\frac{1}{32}-\frac{i\sqrt3}{96}$ &
IYIX & $-\frac{1}{32}+\frac{i\sqrt3}{96}$ &
IYIY & $\frac{\sqrt3}{96}+\frac{i}{32}$ \\

IYZX & $\frac{1}{32}-\frac{i\sqrt3}{96}$ &
IYZY & $-\frac{\sqrt3}{96}-\frac{i}{32}$ &
XIXI & $-\frac{\sqrt3}{96}+\frac{i}{32}$ &
XIXZ & $\frac{\sqrt3}{96}-\frac{i}{32}$ &
XIYI & $\frac{1}{32}+\frac{i\sqrt3}{96}$ &
XIYZ & $-\frac{1}{32}-\frac{i\sqrt3}{96}$ \\

XXXX & $\frac{\sqrt3}{48}$ &
XXXY & $-\frac{i\sqrt3}{48}$ &
XXYX & $-\frac{i\sqrt3}{48}$ &
XXYY & $-\frac{\sqrt3}{48}$ &
XYXX & $-\frac{i\sqrt3}{48}$ &
XYXY & $-\frac{\sqrt3}{48}$ \\

XYYX & $-\frac{\sqrt3}{48}$ &
XYYY & $\frac{i\sqrt3}{48}$ &
XZXI & $\frac{\sqrt3}{96}-\frac{i}{32}$ &
XZXZ & $-\frac{\sqrt3}{96}+\frac{i}{32}$ &
XZYI & $-\frac{1}{32}-\frac{i\sqrt3}{96}$ &
XZYZ & $\frac{1}{32}+\frac{i\sqrt3}{96}$ \\

YIXI & $\frac{1}{32}+\frac{i\sqrt3}{96}$ &
YIXZ & $-\frac{1}{32}-\frac{i\sqrt3}{96}$ &
YIYI & $\frac{\sqrt3}{96}-\frac{i}{32}$ &
YIYZ & $-\frac{\sqrt3}{96}+\frac{i}{32}$ &
YXXX & $-\frac{i\sqrt3}{48}$ &
YXXY & $-\frac{\sqrt3}{48}$ \\

YXYX & $-\frac{\sqrt3}{48}$ &
YXYY & $\frac{i\sqrt3}{48}$ &
YYXX & $-\frac{\sqrt3}{48}$ &
YYXY & $\frac{i\sqrt3}{48}$ &
YYYX & $\frac{i\sqrt3}{48}$ &
YYYY & $\frac{\sqrt3}{48}$ \\

YZXI & $-\frac{1}{32}-\frac{i\sqrt3}{96}$ &
YZXZ & $\frac{1}{32}+\frac{i\sqrt3}{96}$ &
YZYI & $-\frac{\sqrt3}{96}+\frac{i}{32}$ &
YZYZ & $\frac{\sqrt3}{96}-\frac{i}{32}$ &
ZXIX & $\frac{\sqrt3}{96}+\frac{i}{32}$ &
ZXIY & $\frac{1}{32}-\frac{i\sqrt3}{96}$ \\

ZXZX & $-\frac{\sqrt3}{96}-\frac{i}{32}$ &
ZXZY & $-\frac{1}{32}+\frac{i\sqrt3}{96}$ &
ZYIX & $\frac{1}{32}-\frac{i\sqrt3}{96}$ &
ZYIY & $-\frac{\sqrt3}{96}-\frac{i}{32}$ &
ZYZX & $-\frac{1}{32}+\frac{i\sqrt3}{96}$ &
ZYZY & $\frac{\sqrt3}{96}+\frac{i}{32}$ \\
\hline
\end{tabular}
\label{tab:pauli_expansion}
\end{table*}

The braiding statistics in this Abelian model are fully characterized by scalar phase factors. When one anyon is moved around another, the system's state acquires a global phase dependent only on the anyon types. A key example is the braiding of a pure flux $m^h$ and a pure charge $e^g$, which yields a phase factor $\omega^{gh}$, where $\omega = e^{i2\pi/3}$. 

As established in Ref.~\cite{Uniquely2023Hai}, the complete braiding statistics of bulk anyons can be faithfully extracted from half-braiding processes involving boundary excitations. In quantum double models with gapped boundaries, certain bulk anyons are condensed at the edge, and can be locally created and annihilated by boundary operators. Consequently, a process wherein an anyon is created at the boundary, transported along a semicircular path around another excitation, and subsequently annihilated at the boundary results in a final state that differs from the initial state solely by a global phase.

Accessing full braiding statistics via half-braiding significantly reduces the number of qubits required for simulation. It becomes feasible to simulate a minimal instance of the $D(\mathbb{Z}_3)$ model using only two qutrits. This encoding leverages the fact that the relevant Hilbert space for a $\mathbb{Z}_3$ charge or flux degree of freedom is three-dimensional, corresponding to the group elements $\{0,1,2\}$. 

\section{Jackknife Uncertainty of the Experimental Phase}
\label{appendix jackknife analysis}

For the experimental phase estimates in Sec.~\ref{braiding data analysis}, we use delete-one jackknife resampling of the measured data. Here, each data set contains $M=100$ reconstructed states $\rho_s$, corresponding to complete RC sequences or bare-circuit repetitions. The reported phase is
\[
\widehat{\phi}=\phi(\bar\rho),\qquad
\bar\rho=\frac{1}{M}\sum_{s=1}^{M}\rho_s,
\]
where $\phi(\rho)$ is defined by Eq.~(\ref{eq:phi2}). Deleting the $s$-th data unit gives
\[
\bar\rho_{(-s)}=\frac{1}{M-1}\sum_{k\ne s}\rho_k,
\qquad
\widehat{\phi}_{(-s)}=\phi(\bar\rho_{(-s)}).
\]
The jackknife standard error is
\[
\begin{aligned}
\mathrm{SE}_{\mathrm{jack}}
&=\left[\frac{M-1}{M}\sum_{s=1}^{M}
\left(\widehat{\phi}_{(-s)}-\bar\phi_{(-)}\right)^2\right]^{1/2},
\\
\bar\phi_{(-)}&=\frac{1}{M}\sum_{s=1}^{M}\widehat{\phi}_{(-s)}.
\end{aligned}
\]
All phases are evaluated on the same continuous branch around $\widehat{\phi}$. The factor $(M-1)/M$ accounts for the overlapping delete-one samples; these samples are not treated as independent experimental runs. The central value remains $\phi(\bar\rho)$. This uncertainty estimate assumes independent data units and stable measurement conditions. For RC, it includes the finite-shot fluctuations and sequence-to-sequence variation present in the measured states, but does not quantify a common systematic calibration bias.

\section{Perturbative Origin of the Dephasing Wall \label{appendix dephasing wall}}

In this appendix we give a perturbative account of the ``dephasing wall'' discussed in Sec.~\ref{section braiding experiment}.

Consider a single Clifford layer of an $N$-qubit quantum circuit consisting of an ideal unitary $U_0$, a coherent error $U_e = e^{-i\theta\boldsymbol{\Gamma}}$ ($\theta\ll1$, $\boldsymbol{\Gamma}$ a generalized Pauli matrix satisfying $\boldsymbol{\Gamma}^2=I$), and a Pauli dephasing channel $\mathcal{D}$. The Clifford assumption ensures that the compensating operation defined below is also a Pauli operator. The overall evolution without error mitigation is $\mathcal{D}\bigl( U_e U_0 \rho_{\mathrm{in}} U_0^\dagger U_e^\dagger \bigr).$ We insert a random Pauli gate $P$ before the layer and a compensating gate $Q$ after it such that $Q U_0 P = U_0$. The overall evolution becomes $\mathcal{D}\bigl(Q U_e Q^\dagger U_0 \rho_{\mathrm{in}}  U_0^\dagger Q U_e^\dagger Q^\dagger \bigr)$. Therefore
\[
\begin{aligned}
\rho_{\text{RC}} &= \frac{1}{|\mathcal{P}|}\sum_{Q\in\mathcal{P}} \mathcal{D}\bigl(Q U_e Q^\dagger U_0 \rho_{\mathrm{in}}  U_0^\dagger Q U_e^\dagger Q^\dagger \bigr) \\ &= \mathcal{D} \bigl(\frac{1}{|\mathcal{P}|} \sum_{Q\in\mathcal{P}} \bigl(Q U_e Q^\dagger U_0 \rho_{\mathrm{in}}  U_0^\dagger Q U_e^\dagger Q^\dagger \bigr)\bigr).
\end{aligned}
\]
Now expand $U_e$ to second order in $\theta$: $U_e = I - i\theta\boldsymbol{\Gamma} - \frac{\theta^2}{2}I + \mathcal{O}(\theta^3)$. Then $Q U_e Q^\dagger = I - i\theta\, Q\boldsymbol{\Gamma} Q^\dagger - \frac{\theta^2}{2}I + \mathcal{O}(\theta^3).$ Substituting into the twirling average gives
\[
\begin{aligned}
&\frac{1}{|\mathcal{P}|}\sum_{Q} \bigl (Q U_e Q^\dagger U_0 \rho_{\mathrm{in}}  U_0^\dagger Q U_e^\dagger Q^\dagger \bigr ) = U_0 \rho_{\mathrm{in}}  U_0^\dagger \\
&\qquad + i\theta \frac{1}{|\mathcal{P}|}\sum_{Q} \bigl(U_0 \rho_{\mathrm{in}}  U_0^\dagger Q\boldsymbol{\Gamma} Q^\dagger - Q\boldsymbol{\Gamma} Q^\dagger U_0 \rho_{\mathrm{in}}  U_0^\dagger \bigr) \\
&\qquad + \theta^2 \frac{1}{|\mathcal{P}|}\sum_{Q} \bigl( -U_0 \rho_{\mathrm{in}}  U_0^\dagger + Q\boldsymbol{\Gamma} Q^\dagger U_0 \rho_{\mathrm{in}}  U_0^\dagger Q\boldsymbol{\Gamma} Q^\dagger \bigr) \\
&\qquad + \mathcal{O}(\theta^3).
\end{aligned}
\]
\noindent The $N$-qubit Pauli group is a 1‑design; therefore, $\sum_{Q} Q\boldsymbol{\Gamma} Q^\dagger = 0$. Consequently, 
\[
\begin{aligned}
&\frac{1}{|\mathcal{P}|}\sum_{Q} \bigl (Q U_e Q^\dagger U_0 \rho_{\mathrm{in}}  U_0^\dagger Q U_e^\dagger Q^\dagger \bigr ) \\ &= U_0 \rho_{\mathrm{in}}  U_0^\dagger - \theta^2 \bigl(U_0 \rho_{\mathrm{in}}  U_0^\dagger - \overline{\boldsymbol{\Gamma}U_0 \rho_{\mathrm{in}}  U_0^\dagger\boldsymbol{\Gamma}} \bigr) + \mathcal{O}(\theta^3),
\end{aligned}
\]
\noindent where $\overline{\boldsymbol{\Gamma}U_0 \rho_{\mathrm{in}}  U_0^\dagger\boldsymbol{\Gamma}} \equiv \frac{1}{|\mathcal{P}|}\sum_{Q} Q\boldsymbol{\Gamma} Q^\dagger U_0 \rho_{\mathrm{in}}  U_0^\dagger Q\boldsymbol{\Gamma} Q^\dagger.$ Inserting this result into the expression for $\rho_{\text{RC}}$ yields
\[
\begin{aligned}
\rho_{\text{RC}} = &\mathcal{D}(U_0 \rho_{\mathrm{in}}  U_0^\dagger) + \theta^2 \Bigl( \mathcal{D}\bigl(\overline{\boldsymbol{\Gamma}U_0 \rho_{\mathrm{in}}  U_0^\dagger\boldsymbol{\Gamma}}\bigr) - \mathcal{D}(U_0 \rho_{\mathrm{in}}  U_0^\dagger) \Bigr) \\ &+ \mathcal{O}(\theta^3).
\end{aligned}
\]
Therefore, to the first order in $\theta$, we obtain
\[
\rho_{\text{RC}} = \mathcal{D}(U_0 \rho_{\mathrm{in}}  U_0^\dagger) + \mathcal{O}(\theta^2).
\]
Repeating the above reasoning for an $n$-layer circuit with layer-dependent coherent-error strengths $\{\theta_j\}$ yields

\begingroup
\[
\begin{aligned}
\rho_{\text{RC}} = &\mathcal{D}(U_n \mathcal{D}(U_{n-1} \mathcal{D}(\cdots \mathcal{D}(U_1 \rho_{\mathrm{in}} U_1^\dagger) \cdots) U_{n-1}^\dagger)  U_n^\dagger) \\ &+ \mathcal{O}\left(\sum_{j=1}^{n}\theta_j^2\right).
\end{aligned}
\]
\endgroup
\noindent Therefore, through first order in the coherent-error strengths, RC removes the coherent contribution while the Pauli dephasing channel remains intertwined with the ideal evolution. Residual coherent contributions enter at second order. This perturbative result explains the dephasing-dominated residual observed in Fig.~\ref{fig:braiding2} under the assumptions stated above.

Within this model, increasing the number of RC sequences cannot remove the dephasing contribution. Lowering the residual therefore requires suppressing dephasing itself rather than increasing the amount of randomization.

\bibliographystyle{refs}
\begingroup
\let\RCoriginalbibitem\bibitem
\def\RCupdatedreferenceA{beale2018coherence}
\def\RCupdatedreferenceB{Uniquely2023Hai}
\bibliography{refer}
\endgroup

\end{document}